\documentclass[11pt,a4paper]{article}
 \pdfoutput=1
\usepackage{jcappub}
\usepackage{amsmath}
\usepackage{amsfonts,color}
\usepackage{amssymb,float}
\usepackage{mathtools}
\usepackage[utf8]{inputenc}
\usepackage{url}
\usepackage{graphicx}
\usepackage{refstyle}
\usepackage{wasysym}
\usepackage{tabularx}
\usepackage{accents}
\usepackage{graphicx}
\usepackage{color}
\usepackage[dvipsnames]{xcolor}
\usepackage{hyperref}

\usepackage{bm}

\hypersetup{
    colorlinks=true,
    linkcolor=blue,
    filecolor=magenta,      
    citecolor=red
}

\usepackage{cancel} 
\usepackage{subfigure} 
\usepackage{hyperref} 

\makeatletter
\long\def\dddddot#1{%
  {\mathop {#1}\limits ^{\vbox to-1.4\ex@ {\kern -\tw@ \ex@ \hbox {\normalfont .....}\vss }}}%
}
\long\def\multidots#1#2{%
  \count@=0
  {{\mathop {#2}\limits ^{\vbox to-1.4\ex@ {\kern -\tw@ \ex@ \hbox {\normalfont %
  \loop%
  \ifnum#1>\count@%
  .%
  \advance\count@ by1%
  \repeat%
  }\vss }}}}%
}
\makeatother

\usepackage[dvipsnames]{xcolor}
\usepackage[normalem]{ulem} 

\title{\boldmath Pulsar and cosmic variances of pulsar timing-array correlation measurements of the stochastic gravitational wave background}

\author[a,1]{Reginald Christian Bernardo\note{Corresponding Author}}
\author[a,b]{Kin-Wang Ng}

\affiliation[a]{Institute of Physics, Academia Sinica, Taipei 11529, Taiwan}
\affiliation[b]{Institute of Astronomy and Astrophysics, Academia Sinica, Taipei 11529, Taiwan}

\emailAdd{rbernardo@gate.sinica.edu.tw}
\emailAdd{nkw@phys.sinica.edu.tw}

\abstract{
\textbf{
Pulsar timing-array correlation measurements offer an exciting opportunity to test the nature of gravity in the cosmologically novel nanohertz gravitational wave regime. The stochastic gravitational wave background is assumed Gaussian and random, while there are limited pulsar pairs in the sky. This brings theoretical uncertainties to the correlation measurements, namely the pulsar variance due to pulsar samplings and the cosmic variance due to Gaussian signals.  We demonstrate a straightforward calculation of the mean and the variances on the Hellings-Downs correlation relying on a power spectrum formalism. We keep arbitrary pulsar distances and consider gravitational wave modes beyond Einstein gravity as well as off the light cone throughout, thereby presenting the most general and, most importantly, numerically efficient calculation of the variances.
}}

\begin{document}

\maketitle
\flushbottom


\section{Introduction}
\label{sec:intro}

The direct observation of gravitational wave (GW) is revolutionary in so many ways \cite{LIGOScientific:2016aoc}. For one, it confirms the hundred year speculation about GWs physical existence, and cements general relativity (GR) as one of the most successful scientific theories \cite{LIGOScientific:2017vwq, LIGOScientific:2021sio}. In terms of opening up science prospects, GWs are plausibly the best way to understand the origin of the Universe, as it probes beyond the cosmic microwave background (CMB), astronomical observations brick wall, utilizing gravity's relatively weak coupling with matter to its advantage. In this direction, as with electromagnetic waves, different GW frequency bands tie in to different astrophysical sources, which correspond to specific epochs in the cosmic history. The ground based detectors \cite{LIGOScientific:2021djp, KAGRA:2018plz} for instance are sensitive to frequencies $10^0 - 10^3$ hertz, while planned space based detectors \cite{LISA:2022kgy, TianQin:2015yph}, free from terrestrial restrictions, are expected to be sensitive to frequencies $10^{-4} - 10^{-1}$ hertz. These detectors target mostly compact sources such as Solar mass black holes and neutron stars in a binary, intermediate mass ratio binaries, and extreme mass ratio binaries, that paint the gravitational picture of the late Universe.

A rather innovative means of directly observing GWs is by a pulsar timing array (PTA) \cite{Detweiler:1979wn, Burke-Spolaor:2018bvk}. In this manner, the stochastic gravitational wave background (SGWB) correlates the time of arrival of radio pulses of pulsars with one another, which leaves a distinct signal known as the Hellings-Downs (HD) curve \cite{Hellings:1983fr}. In contrast with ground and space based GW observatories, in a PTA, the distance from the earth to the pulsars serves as the GW detector arms, permitting access to the nanohertz GW band, targeting sources that bring in invaluable information about the early cosmic evolution, such as phase transitions in the early Universe, cosmic strings, and supermassive binary black holes \cite{Romano:2019yrj}. Current PTAs are the North American Nanohertz Observatory for Gravitational Waves (NANOGrav) \cite{NANOGrav:2020bcs}, the Parkes Pulsar Timing Array (PPTA) \cite{Shannon:2015ect}, the European Pulsar Timing Array (EPTA) \cite{Lentati:2015qwp}, which together form the International Pulsar Timing Array (IPTA) \cite{2010CQGra..27h4013H}. These efforts collectively have been observing about a hundred millisecond pulsars for years now in hopes of providing evidence of the SGWB, and their current data do not disappoint, featuring a common spectrum process across the millisecond pulsars in the PTA, consistent with predictions from GR for several potential GW sources. However, the noise in the present PTA data sets are quite large to be able to infer whether the spatial correlations are due to SGWB. This may be due to the limitations in the optimal statistic analysis, among others, but recently the question of whether variances in the Hellings-Downs correlation also play a role has been put forward \cite{Allen:2022dzg, Allen:2022ksj}. Simply put, the variances also bring in information about the nature of the sources, and must be given attention.

In this paper, we show how the theoretical uncertainties in the GW correlation measurements in a PTA may be studied alternatively using a power spectrum formalism (PSF) \cite{Qin:2018yhy, Qin:2020hfy, Ng:2021waj, Liu:2022skj}. Taking in historical lessons learned from the CMB, we resort to the multipoles of the correlation in obtaining the PTA observables. This provides an elegant path to calculate not just the mean but the higher moments of an observable, as we effortlessly demonstrate by reproducing the variances of the Hellings-Downs correlation. On top of that, this paves the road to the calculation of the theoretical uncertainties of the general subluminal tensor, vector, and scalar metric polarizations that are important for recognizing alternative gravity degrees of freedom that may be present in this uncharted nanohertz GW cosmic territory.

The IPTA observes a hundred pulsars. It suffices to use correlation functions to perform data analysis. As PTA science comes of age with a few thousands of pulsars, inevitably, the analysis will naturally rely on the PSF. This is important in order to extract precise scientific information about the sources of the SGWB, and eventually in measuring anisotropies. The process then boils down to a calculation of the power spectrum of the SGWB. In \cite{Bernardo:2022rif}, we presented a simple recipe how this could be done for any GW polarizations, at any speeds, and with arbitrary pulsar distances. This can be coded with ease in any programming language, e.g., python \cite{10.5555/1593511} or Julia \cite{Julia-2017}, which may later anchor data analysis routines for SGWB detection in PTA observations. Here, we use the resulting multipoles to calculate the pulsar and cosmic variances in the overlap reduction function (ORF), PTA's main spatial correlation observable. 

This paper proceeds as follows. In Section \ref{sec:pulsar_timing}, we briefly review pulsar timing and introduce the correlation operator. In Section \ref{sec:pta_variances_psmeth}, we present the PSF for calculating the mean and the variance. Then, this is utilized in Section \ref{sec:hdmeanvar} to look at the uncertainty of the HD correlation with pulsar timing array observation. The rest of the paper then deals with a showcase of the variances for subluminal tensor (Section \ref{sec:tensorpols}), vector (Section \ref{sec:vectorpols}), and scalar (Section \ref{sec:scalarpols}) metric polarizations, demonstrating the advantage of the PSF in effortlessly ushering in the most general and numerically efficient way of calculating pulsar timing array observables. We draw some final remarks in Section \ref{sec:discussion}.

\section{Pulsar timing}
\label{sec:pulsar_timing}

We briefly review the pulsar timing residual (Section \ref{subsec:timingres}) and discuss the correlation operator (Section \ref{subsec:corrop}).

\subsection{Timing residual}
\label{subsec:timingres}

A PTA's direct observable of the individual pulsars is the timing residual, $r\left(t, \hat{e}\right)$, which can be written as an integral over the GW induced redshift fluctuation, $z\left(\eta, \hat{e}\right)$, as 
\begin{equation}
\label{eq:timing_residual}
    r\left(t, \hat{e}\right) = \int_0^t dt' \ z\left(t', \hat{e}\right) \,,
\end{equation}
where $t$ corresponds to the duration of the observation and $\hat{e}$ is a unit vector pointing toward the pulsar from earth. The redshift fluctuation on one hand can be derived from a Sachs-Wolfe integral,
\begin{equation}
\label{eq:z_swolf}
    z\left(t', \hat{e}\right) = - \dfrac{1}{2} \int_{t' + \eta_e}^{t' + \eta_r} d\eta \ d^{ij} \partial_\eta h_{ij} \left( \eta, \vec{x} \right) \,,
\end{equation}
where the detector tensor is given by
\begin{equation}
    d^{ij} = \hat{e}^i \otimes \hat{e}^j \,.
\end{equation}
Let us write down the GW as a superposition of various polarizations $A$, frequencies $f$, and propagation directions $\hat{k}$ and speeds $v$,
\begin{equation}
\label{eq:gw_general}
    h_{ij}\left(\eta, \vec{x}\right) = \sum_A \int_{-\infty}^\infty df \int_{S^2} d\hat{k} \ h_A\left(f, \hat{k}\right) \varepsilon_{ij}^A e^{-2\pi i f \left( \eta - v \hat{k} \cdot \vec{x} \right)} \,,
\end{equation}
that satisfies the stochastic and Gaussian property for an isotropic SGWB,
\begin{equation}
    \langle h_A \left( f, \hat{k} \right) h^*_{A'} \left(f', \hat{k}'\right) \rangle= P_{AA'}\left(f\right)\delta_{AA'} \delta \left(f - f'\right) \delta \left( \hat{k} - \hat{k}' \right) \,,
\end{equation}
where $P_{AA}(f)$ is the SGWB power spectrum. Then, 
we are able to express the two point function of the timing residual for a pair of pulsars $a$ and $b$ as \cite{Bernardo:2022rif}
\begin{equation}
\label{eq:timingtwopoint}
    \langle r\left(t_a, \hat{e}_a\right) r\left(t_b, \hat{e}_b\right) \rangle = \sum_A \int_{-\infty}^\infty df \ \left( 1 - e^{-2\pi i f t_a} \right) \left( 1 - e^{2\pi i f t_b} \right) \dfrac{P_{AA}\left(f\right) }{2\pi^{3/2}f^2} \gamma_{ab}^A \left( \hat{e}_a \cdot \hat{e}_b \right) \,,
\end{equation}
where $\gamma_{ab}^A \left( \hat{e}_a \cdot \hat{e}_b \right)$ is the ORF for a GW with polarization $A$. It is a function dependent on frequency $f$, speed $v$, and pulsar distances $D_a$ and $D_b$.
Later in the next section, (\ref{eq:timingtwopoint}) will relate the timing residual two point correlation coefficients, $D_l$, with the ORF's power spectrum multipoles, $C_l$.

In what follows, we suppress the superscript $A$ standing for the GW polarizations and introduce the shorthand for the timing residual, ${\bm r}_a = r\left(t_a, \hat{e}_a\right)$, for brevity. Needless to say, the subsequent mathematical results obtained using the PSF hold for any GW polarization. Furthermore, we assume that $t_a = t_b$, $D_a = D_b$, and a narrow power spectrum $P_{AA}(f)$. These assumptions are indeed pertinent to realistic observation, such as the SGWB generated by subhorizon processes \cite{Qin:2020hfy, Ng:2021waj}.

Moving on, we write down the timing residual, now in shorthand notation, as a Laplace series,
\begin{equation}
\label{eq:timingres_harmonicseries}
    {\bm r}_a = \sum_{lm} {\bm a}_{lm} Y_{lm} \left( \hat{e}_a \right) \,,
\end{equation}
where the $Y_{lm}(\hat{e})$'s are the spherical harmonics.
Assuming spatial isotropy, or rather in terms of the timing residual's multipoles,
\begin{equation}
    \langle {\bm a}_{lm}^* {\bm a}_{l'm'} \rangle = D_l \delta_{ll'} \delta_{mm'} \,,
\end{equation}
where $D_l$ quantifies the two point timing residual power spectrum. Calculating the two point function, we therefore obtain
\begin{equation}
    S_{ab}\equiv \langle {\bm r}_a {\bm r}_b \rangle = \sum_l \dfrac{2l + 1}{4\pi} D_l P_l \left( \hat{e}_a \cdot \hat{e}_b \right) \,, 
\end{equation}
after considering the completeness relation
\begin{equation}
\label{eq:Ycomplete}
    P_l \left( \hat{e}_a \cdot \hat{e}_b \right) = \dfrac{4\pi}{2l + 1} \sum_m Y^*_{l m} \left( \hat{e}_a \right) Y_{l m} \left( \hat{e}_b \right) \,,
\end{equation}
where $P_l(x)$'s are the Legendre polynomials. 

Meanwhile, we look at a few more identities linking the timing residual to other physical quantities. Considering only Gaussian fields, and thus accommodating the factorization of the four point function as
\begin{equation}
    \langle {\bm r}_a {\bm r}_b {\bm r}_c {\bm r}_d \rangle = \langle {\bm r}_a {\bm r}_b \rangle \langle {\bm r}_c {\bm r}_d \rangle +
    \langle {\bm r}_a {\bm r}_c \rangle \langle {\bm r}_b {\bm r}_d \rangle + \langle {\bm r}_b {\bm r}_c \rangle \langle {\bm r}_a {\bm r}_d \rangle \,,
\end{equation}
we obtain the Wick rotation
\begin{equation}
\label{eq:a_wickrotation}
\begin{split}
    \langle {\bm a}^*_{l_1 m_1} {\bm a}_{l_2 m_2} {\bm a}^*_{l_3 m_3} {\bm a}_{l_4 m_4} \rangle = \ & D_{l_1} D_{l_3} \delta_{l_1 l_2} \delta_{m_1 m_2} \delta_{l_3 l_4} \delta_{m_3 m_4} + D_{l_1} D_{l_2} \delta_{l_1 l_4} \delta_{m_1 m_4} \delta_{l_2 l_3} \delta_{m_2 m_3} \\
    & + (-1)^{m_1} (-1)^{m_2} D_{l_1} D_{l_2} \delta_{l_1 l_3} \delta_{m_1 -m_3} \delta_{l_2 l_4} \delta_{m_2 -m_4} \,.
\end{split}
\end{equation}
With this, we can calculate the variance of our estimator of ${\bm r}_a {\bm r}_b$:
\begin{equation}
    \left( \Delta S_{ab} \right)^2 = \langle ({\bm r}_a {\bm r}_b)^2 \rangle - \langle {\bm r}_a {\bm r}_b \rangle^2 \,.
\end{equation}
The second moment of ${\bm r}_a {\bm r}_b$ turns out to be
\begin{equation}
\label{eq:timingsecondmoment}
\begin{split}
    \langle \left( {\bm r}_a {\bm r}_b \right)^2 \rangle = 2 \left( \sum_{l m} D_l Y^*_{lm}\left(\hat{e}_a\right)Y_{lm}\left(\hat{e}_b\right) \right)^2 + \left( \sum_{l m} D_l Y^*_{lm}\left(\hat{e}_a\right)Y_{lm}\left(\hat{e}_a\right) \right)^2 \,.
\end{split}
\end{equation}
Recognizing the first term as two times the square of the estimator $\langle {\bm r}_a {\bm r}_b \rangle$, we obtain the variance
\begin{equation}
    \left( \Delta S_{ab} \right)^2 = \left( \sum_l \dfrac{2l + 1}{4\pi} D_l P_l \left( \hat{e}_a \cdot \hat{e}_b \right) \right)^2 + \left( \sum_l \dfrac{2l + 1}{4\pi} D_l P_l \left( 0 \right) \right)^2 \,.
\end{equation}
We kept the details of the derivation above to a minimum, but later in the next section we show a similar calculation when obtaining the variance of the ORF.

\subsection{Correlation operator}
\label{subsec:corrop}

We largely identify a GW correlation via the operator,
\begin{equation}
\label{eq:cop}
    {\bm\gamma}_{ab} = {\bm\beta}_a^\dagger {\bm\beta}_b \,,
\end{equation}
hereinafter referred to as the `correlation operator' where the quantity ${\bm \beta}_a$ admits the multipolar expansion given by
\begin{equation}
\label{eq:beta_hseries}
    {\bm \beta}_a = \sum_{lm} {\bm b}_{lm} Y_{lm}\left(\hat{e}_a\right) \,.
\end{equation}
The quantities ${\bm \gamma}_{ab}$ and ${\bm \beta}_a$ represent the SGWB induced spatial correlation, $\langle {\bm r}_a {\bm r}_b \rangle$, and the timing residue, ${\bm r}_a$, i.e., they are proportional up to constants, and so carry their physical meaning. We shall see how these multipolar coefficients, ${\bm b}_{lm}$, relate to the timing residual multipoles, ${\bm a}_{lm}$, that is after we first setup its connection with the power spectrum.

The ORF can be identified by the \textit{ensemble} average, denoted by $\langle \cdots \rangle$, of the correlation operator. To see this, we substitute (\ref{eq:beta_hseries}) into (\ref{eq:cop}), and take the ensemble average to obtain
\begin{equation}
    \langle {\bm\beta}_a^\dagger {\bm\beta}_b \rangle = \sum_{l_1 m_1} \sum_{l_2 m_2} \langle {\bm b}^\dagger_{l_1 m_1} {\bm b}_{l_2 m_2} \rangle Y^*_{l_1 m_1} \left( \hat{e}_a \right) Y_{l_2 m_2} \left( \hat{e}_b \right) \,.
\end{equation}
Then, assuming spatial isotropy, we write down 
\begin{equation}
\label{eq:isotropic}
    \langle {\bm b}^\dagger_{l_1 m_1} {\bm b}_{l_2 m_2} \rangle = C_{l_1} \delta_{l_1 l_2} \delta_{m_1 m_2} \,,
\end{equation}
where the $C_l$'s are the power spectrum multipoles of the SGWB \cite{Bernardo:2022rif}. The ensemble average of the correlation operator reduces to
\begin{equation}
    \langle {\bm\beta}_a^\dagger {\bm\beta}_b \rangle = \sum_{l m} C_l Y^*_{l m} \left( \hat{e}_a \right) Y_{l m} \left( \hat{e}_b \right) \,.
\end{equation}
Utilizing the completeness identity of the spherical harmonics (\ref{eq:Ycomplete}), we are further able to obtain \cite{Bernardo:2022rif}
\begin{equation}
    \langle {\bm\beta}_a^\dagger {\bm\beta}_b \rangle = \sum_{l} \dfrac{2l + 1}{4\pi} C_l P_l \left( \hat{e}_a \cdot \hat{e}_b \right) \,,
\end{equation}
which is the well known expression of the ORF given the angular power spectrum multipoles $C_l$ \cite{Qin:2018yhy, Qin:2020hfy, Ng:2021waj, Liu:2022skj, Bernardo:2022rif}.
This permits the identification of the ORF with the correlation operator as
\begin{equation}
    \gamma_{ab}\left(\zeta\right) = \langle {\bm\beta}_a^\dagger {\bm\beta}_b \rangle \,.
\end{equation}

We ask if there is a relation between the correlation operator and the timing residual. To answer this simply, we look at (\ref{eq:timingtwopoint}) considering a fixed GW polarization and frequency. Under the assumptions, $t_a = t_b= t$, $D_a = D_b =D$, a narrow power spectrum $P_{AA}(f)$, considering only Gaussian fields, we obtain the relation
\begin{equation}
    D_l =  \left\vert 1 - e^{-2\pi i f t} \right\vert^2 \dfrac{P_{AA}\left(f\right) }{2\pi^{3/2}f^2} C_l \,,
\end{equation}
between the timing residual's two point correlation coefficients and the power spectrum multipoles of the SGWB correlation. 
For a general GW mixture of various polarizations and frequencies, this result naturally generalizes by summing and integrating the right hand side over polarizations and frequencies. We always find $D_l = \text{Constant} \times C_l$ for some constant. Thus, ${\bm b}_{lm}$ inherits the statistics (\ref{eq:a_wickrotation}) from ${\bm a}_{lm}$, and that the variance in ${\bm r}_a {\bm r}_b$ is equal to the variance in ${\bm \beta}_a {\bm \beta}_b$, up to an overall factor.

The HD correlation may be obtained for the luminal tensor induced GW correlation in the infinite distance limit. In symbols,
\begin{equation}
\label{eq:hdlim}
    {\bm \gamma}_{ab}^\text{HD} = {\bm \gamma}_{ab}^\text{T}|_{v \rightarrow 1, D\rightarrow \infty} \,,
\end{equation}
where $D$ stands for the pulsars' distances from the observer. The HD curve comes out as the ensemble average of the HD operator,
\begin{equation}
    \gamma_{ab}^\text{HD} \left( \zeta \right) = \langle {\bm \gamma}_{ab}^\text{HD} \rangle \,,
\end{equation}
where $\zeta$ is the angular separation of a pulsar pair, that is, $\cos \zeta = \hat{e}_a \cdot \hat{e}_b$ between pulsars $a$ and $b$.

In the following section, we obtain the variances of the correlation operator which relates to the theoretical uncertainties. We also compute the variance in the power spectrum of the SGWB correlation multipoles.

\section{ORF variances through the power spectrum}
\label{sec:pta_variances_psmeth}

We present the PSF by calculating the mean and the variance of a GW correlation in a pulsar timing array measurement. Utilizing the correlation operator (Section \ref{subsec:corrop}) and then calculating the total variance (Section \ref{subsec:totalvar}) and the cosmic variance (Section \ref{subsec:cosmicvar}).

\subsection{Total variance}
\label{subsec:totalvar}

The total variance is the variance of a single pulsar pair, whose pular timing residuals are correlated by the SGWB. We calculate this in terms of the power spectrum multipoles below.

Now, we want to calculate the variance, that is,
\begin{equation}
\label{eq:variance}
    \left(\Delta \gamma_{ab} \right)^2 = \langle \left( {\bm \beta}_a^\dagger \bm{\beta}_b \right)^2 \rangle - \langle {\bm \beta}_a^\dagger \bm{\beta}_b \rangle^2 \,.
\end{equation}
The second term on the right is the square of the ORF. We thus need to take care of only the first term, which is the second moment of the correlation operator. Thus, the second moment of the correlation operator becomes
\begin{equation}
\label{eq:secondmoment_correlation}
    \langle \left( {\bm \beta}_a^\dagger \bm{\beta}_b \right)^2 \rangle = 2 \left( \gamma_{ab}\left(\zeta\right) \right)^2 + \left( \gamma_{aa} \right)^2 \,,
\end{equation}
where $\gamma_{aa}$ is the autocorrelation. We carry out the detailed steps in Appendix \ref{subsec:totalvar_appendix}. The variance (\ref{eq:variance}) of the correlation is finally
\begin{equation}
\label{eq:totalvariance}
    \left(\Delta \gamma_{ab} \right)^2 = \left( \gamma_{ab}\left(\zeta\right) \right)^2 + \left( \gamma_{aa} \right)^2 \,.
\end{equation}

This total variance (\ref{eq:totalvariance}) stands for the uncertainty in the correlation expected in one pulsar pair. In the HD limit, it is also easy to see that this agrees with the result of \cite{Allen:2022dzg}, referring to the total variance as unpolarized confusion noise. We shall show this explicitly in the next section. But before that, we move to the cosmic variance of the correlation.

\subsection{Cosmic variance}
\label{subsec:cosmicvar}

The cosmic variance comes out of pulsar pairs of a fixed angular separation in the sky. In other words, we perform a full sky averaging over pulsar pairs of the same angular separation. We derive this explicitly using the power spectrum formalism below.

To calculate the cosmic variance, we instead perform a full sky averaging with a fixed angle over a pulsar pair. In symbols, we write this as
\begin{equation}
    \{ \cdots \}_\text{S} = \text{full sky averaging} = \int d\Omega d\Omega' \cdots d\Omega'' \left( \cdots \right) \,.
\end{equation}
The two point spherical harmonics can then be identified as \cite{1998IJMPD...7...89N, Ng:1997ez}
\begin{equation}
    \{ Y_{l'm'}^*\left(\hat{n}'\right) Y_{lm}\left(\hat{n}\right) \}_\text{S} = P_l \left( \cos \zeta \right) \dfrac{\delta_{ll'} \delta_{mm'}}{4\pi} \,,
\end{equation}
where $\zeta$ corresponds to the fixed separation angle on the sky.

To introduce full sky averaging, we start with the first moment of the correlation operator,
\begin{equation}
\begin{split}
    \{ {\bm \beta}_a^\dagger \bm{\beta}_b \}_\text{S} = & \sum_{l_1 m_1} \sum_{l_2 m_2} {\bm b}_{l_1m_1}^\dagger {\bm b}_{l_2m_2} \{ Y_{l_1m_!}^*\left( \hat{e}_a \right) Y_{l_2m_2}\left(\hat{e}_b\right) \}_\text{S} \\
    = & \sum_{l_1 m_1} \sum_{l_2 m_2} {\bm b}_{l_1m_1}^\dagger {\bm b}_{l_2m_2} \left( P_{l_1} \left( \cos \zeta \right) \dfrac{\delta_{l_1l_2} \delta_{m_1m_2}}{4\pi} \right) \\
    \{ {\bm \beta}_a^\dagger \bm{\beta}_b \}_\text{S} = &
    \sum_{lm} \dfrac{ {\bm b}_{lm}^\dagger {\bm b}_{lm} }{4\pi} P_l \left( \cos\zeta \right) \,.
\end{split}
\end{equation}
To simplify this further, we define the operator
\begin{equation}
\label{eq:ps_operator}
    {\bm C}_l = \sum_m \dfrac{{\bm b}_{lm}^\dagger {\bm b}_{lm}}{2l + 1} \,,
\end{equation}
which is related to the power spectrum multipoles (\ref{eq:isotropic}). To see this explicit relation, we perform ensemble averaging over this operator,
\begin{equation}
\begin{split}
    \langle {\bm C}_l \rangle = \ & \sum_m \dfrac{ \langle {\bm b}_{lm}^\dagger {\bm b}_{lm} \rangle }{2l + 1} \\
    = \ & \sum_m \dfrac{ C_l }{2l + 1} \\
    \langle {\bm C}_l \rangle = \ & (2l + 1) \dfrac{ C_l }{2l + 1} \,,
\end{split}
\end{equation}
which leads to
\begin{equation}
\label{eq:psop_ensemble}
    \langle {\bm C}_l \rangle =  C_l \,.
\end{equation}

We then obtain the full sky averaged first moment of the correlation operator as
\begin{equation}
\label{eq:fsave_hd}
    \{ {\bm \beta}_a^\dagger \bm{\beta}_b \}_\text{S} =
    \sum_{l} \dfrac{ 2l + 1 }{4\pi} {\bm C}_l P_l \left( \cos\zeta \right) \,.
\end{equation}
Clearly, this is related to the ORF via an ensemble average,
\begin{equation}
    \langle \{ {\bm \beta}_a^\dagger \bm{\beta}_b \}_\text{S} \rangle = \gamma_{ab}\left(\zeta\right) \,.
\end{equation}

The cosmic variance can be obtained from the full sky averaged second moment of the correlation. In symbols, to obtain the cosmic variance, we calculate
\begin{equation}
\label{eq:cv_fullskydef}
    \text{CV} = \langle \{ {\bm \beta}_a^\dagger \bm{\beta}_b \}_\text{S}^2 \rangle - \langle \{ {\bm \beta}_a^\dagger \bm{\beta}_b \}_\text{S} \rangle ^2 \,.
\end{equation}
The second term is simply the square of (\ref{eq:fsave_hd}), which is one ensemble average away from the ORF. Thus we focus on the first term. We eventually end up with
\begin{equation}
\label{eq:secondmoment_corr_skyave}
    \langle \{ {\bm \beta}_a^\dagger {\bm \beta}_b \}_\text{S}^2 \rangle = \left( \sum_{l} \dfrac{2l + 1}{4\pi} C_l P_l \left( \cos \zeta \right) \right)^2 + \sum_l \dfrac{2l + 1}{8\pi^2} C_l^2 P_l\left(\cos\zeta\right)^2 \,.
\end{equation}
We provide some of the technical steps in Appendix \ref{subsec:cosmicvar_appendix}. The first squared sum term above may be recognized to be $\langle \{ {\bm \beta}_a^\dagger \bm{\beta}_b \}_\text{S} \rangle ^2 = \gamma_{ab}(\zeta)^2$ which is the square of the ORF. Putting all the above information together back into (\ref{eq:cv_fullskydef}), we finally get to the cosmic variance given by
\begin{equation}
\label{eq:cv_ps}
\text{CV} = \sum_l \dfrac{2l + 1}{8\pi^2} C_l^2 P_l\left(\cos\zeta\right)^2 \,.
\end{equation}

This is the uncertainty emerging from having a sufficiently large number of pulsar pair correlation measurements at the same angle. In the HD limit, this agrees with \cite{Allen:2022dzg}. We reveal this agreement explicitly in the next section.

\subsection{Variance in the power spectrum}
\label{subsec:psvar}

We have by far been concerned with the variances in the spatial correlation operator. In this subsection, we instead look at the variance of the power spectrum multipoles.

We recall that the ensemble average of the multipole operator (\ref{eq:ps_operator}) are the power spectrum multipoles (\ref{eq:psop_ensemble}). We now want to compute the variance of the multipole operator, or simply the variance of the multipoles, that is,
\begin{equation}
\label{eq:psop_variance_def}
    \left(\Delta C_l\right)^2 = \langle {\bm C}_l^2 \rangle - \langle {\bm C}_l \rangle^2 \,.
\end{equation}
From this, and (\ref{eq:psop_ensemble}), we obtain the variance of the power spectrum multipoles, $\left(\Delta C_l\right)^2$, to be given by
\begin{equation}
\label{eq:psop_variance}
    \left(\Delta C_l\right)^2 = {2 C_l^2}/(2l + 1)\,,
\end{equation}
or in terms of the uncertainty,
\begin{equation}
    \dfrac{\Delta C_l}{C_l} = \sqrt{\dfrac{2}{2l + 1}} \,.
\end{equation}
Detailed calculation of the ensemble average of the second moment of the power spectrum multipoles, which leads to (\ref{eq:psop_variance}), is given in Appendix \ref{subsec:ps_variance_appendix}.

It is useful to note that this is the same expression as with the temperature anisotropies of the CMB. This shows that a better angular resolution allows measurements of the power spectrum up to a larger $l$, for example, a $\Delta \zeta  = 0.1^\circ$ resolution permits the measurement of the first few thousand power spectrum multipoles, $l \leq 180^\circ/\Delta \zeta \sim 1800$, where the variance can be as small as $\Delta C_l/C_l \sim 1/\sqrt{1800} \sim 1/42$. Understandably, this level of resolution seems to be on the horizon of a PTA given the present measurements, but it would be quite impressive to get here.

\section{Hellings-Downs: mean and variance}
\label{sec:hdmeanvar}

We present the mean and variances of the HD correlation together with the 12.5 year NANOGrav data set \cite{NANOGrav:2020bcs}. This serves as the baseline of the general GW correlation discussion to follow (Sections \ref{sec:tensorpols}, \ref{sec:vectorpols}, and \ref{sec:scalarpols}).

We recall that the power spectrum multipoles of the HD correlation can be shown to be \cite{Gair:2014rwa, Ng:2021waj, Liu:2022skj, Bernardo:2022rif}
\begin{equation}
    C_l^\text{HD} = \dfrac{8 \pi^{3/2}}{(l - 1)l(l + 1)(l + 2)} \,.
\end{equation}
We mention that the above multipoles lead to the ORF, $\gamma_{ab}^\text{HD}\left(\hat{e}_a \cdot \hat{e}_b\right)$, which is related to the normalized ORF conventionally used for data analysis as $\Gamma_{ab}^\text{HD}\left(\hat{e}_a \cdot \hat{e}_b\right) = \gamma_{ab}^\text{HD}\left(\hat{e}_a \cdot \hat{e}_b\right) \times 0.5/\gamma_{ab}^\text{HD}\left(0^+\right)$ such that $\Gamma_{ab}^\text{HD}\left(0^+\right) = 0.5$ \cite{Bernardo:2022rif}.

\begin{figure}[h!]
\center
\includegraphics[width = 0.5 \textwidth]{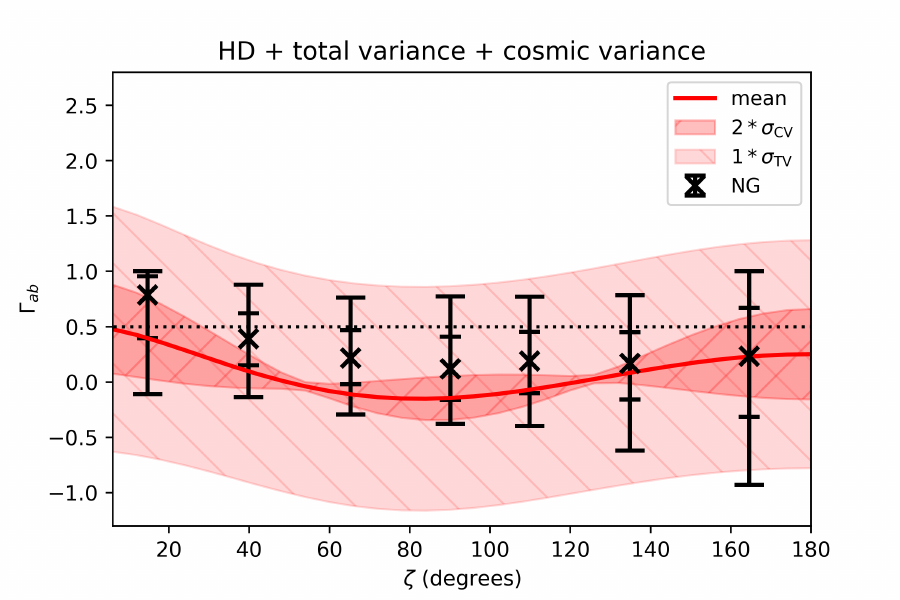}
\caption{(a) Mean and uncertainty of the Hellings-Downs curve from the total (\ref{eq:totalvariance}) and cosmic (\ref{eq:cv_ps}) variances. The $2*\sigma$ error bars and extreme values are obtained from the NANOGrav 12.5 year data \cite{NANOGrav:2020bcs}. We utilize only the first thirty multipoles for the power spectrum calculation. The horizontal dotted line corresponds to a `monopolar' correlation.}
\label{fig:hdtotalvar}
\end{figure}

Figure \ref{fig:hdtotalvar} shows the ORF of the HD curve with its uncertainty emerging from the total and cosmic variances obtained using the power spectrum method. The horizontal dotted line corresponds to a monopolar spatial correlation, shown as a reference, since while a nonGW effect, this is a systematic error that must be properly taken care of in a PTA. Because the HD ORF is normalized as $\Gamma_{ab}^{\rm HD}\left(0\right) = 0.5$, this additionally acts as a visual reference when the correlations are stronger or weaker compared to the HD at small angles.

The total variance $\sigma_\text{TV}^2$ (Figure \ref{fig:hdtotalvar}, red `\textbackslash' hatched region), again, is what we would expect for the uncertainty for a single pulsar pair whose timing residuals are correlated by the luminal transverse tensor SGWB, expected in GR. Propagating this uncertainty to the average cross correlated power in a PTA, we also see that the present NANOGrav data set is consistent with the total variance as an upper bound to its spatial correlation data points. There are 990 pulsar pairs (over 45 pulsars) in the NANOGrav 12.5 year data set. Should the pulsar pairs be uncorrelated, the single pair uncertainty $\Delta \gamma_{ab}^2$ may then be simply propagated as $\Delta \gamma_{ab}^2/N$ for $N$ pulsar pairs in an angular bin. With this in mind, we may take the observation to suggest that there are more than a few pulsar pairs in the current data as the measurement uncertainty is narrower compared to the total variance.

The PTA pulsar pairs are of course also correlated with one another and this cross correlation sustains a generally nonvanishing uncertainty even for an arbitrarily large data set. This takes us to the cosmic variance $\sigma_\text{CV}^2$ of the HD correlation (Figure \ref{fig:hdtotalvar}, red `/' hatched region). As alluded previously, the cosmic variance is what is retained when there is a large number of cross correlated pulsar pairs. This is generally nonvanishing, but notice that it reaches a minimum at certain angular separations, where the mean of the HD correlation hits the zero mark. Near these angles, the data points may be considered as an indication to think about alternative viable descriptions of the SGWB. In Figure \ref{fig:hdtotalvar}, these spatial minima of the cosmic variance appear at $\zeta \sim 55^\circ$ and $\zeta \sim 125^\circ$. Clearly the HD curve and the cosmic variance still falls consistently within the 95\% confidence interval of the observation \cite{NANOGrav:2020bcs}. The current data set however seems quite too conservative with its error bars compared with the HD curve and its cosmic variance, or rather that the error bars seem too large compared to the cosmic variance, given that there are a number of available pulsar pairs that should reduce the total variance. Clearly a more stringent observation would be preferred, in order to make science inference, which is targeted by future PTA missions.

\begin{figure}[h!]
\center
\includegraphics[width = 0.5 \textwidth]{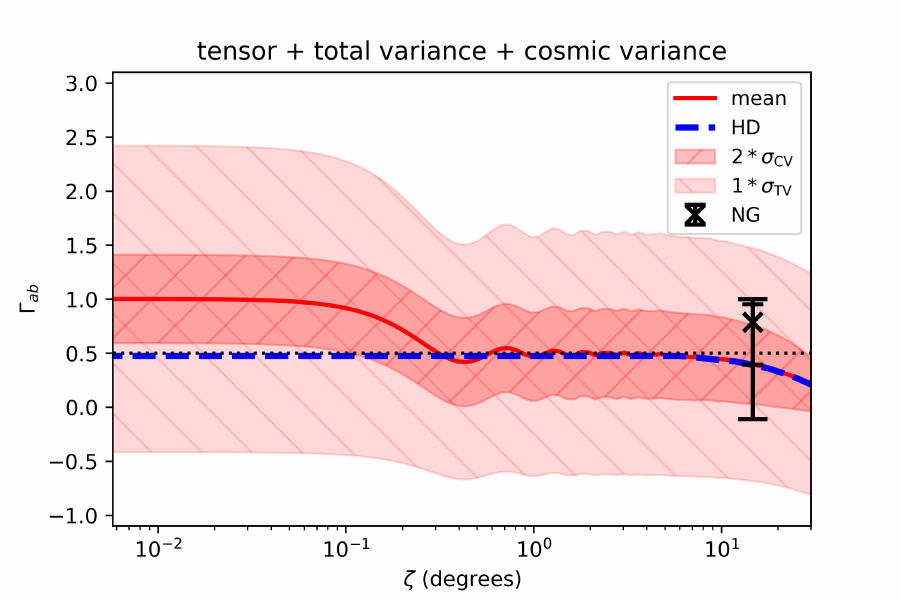}
\caption{The HD ORF at low angles ($\zeta < 30^\circ$) obtained using the PSF. We consider one thousand luminal tensor multipoles $l \leq 1000$ for numerical convergence at such small angles. The horizontal dotted line corresponds to a `monopolar' correlation.}
\label{fig:hdlowangle}
\end{figure}

To see where we might find departures, we look at the low angles of the ORF, demanding at least a thousand multipoles for the calculation. In this case, we take a general luminal tensor mode keeping the pulsars to be at a finite distance \cite{Bernardo:2022rif}, in particular, at $fD = 100$. The result is shown in \ref{fig:hdlowangle} for angles smaller than thirty degrees.

This demonstrates the advantage of the PSF, that it captures the small scale power that is contained by pulsar pairs of a few subdegree separations. We emphasize that this power is otherwise missed by the HD curve, which was obtained by taking the pulsars' to be infinitely distant from the observer, or alternatively by neglecting the so called `pulsar term' as a price for obtaining an analytical expression. Such small angles are a clear place where one might find departures from the HD, which is even two cosmic variances away from the luminal tensor ORF's mean.

We note that Figure \ref{fig:hdlowangle} is presented with the lowest angle data point in NANOGrav's 12.5 year data set \cite{NANOGrav:2020bcs}. This adds emphasis where significant deviations from the canonical description of the nanohertz GW sky may lie. Nontensorial metric polarizations present even a wider range of behavior in this observational region that can distinguish the nature of the gravitational degrees of freedom dominant in the SGWB \cite{Bernardo:2022rif}. The observation of pulsar pairs of a few subdegree separations is clearly something to look forward to in this regard.

We mention that the analytical expressions derived in \cite{Allen:2022dzg} were normalized as $\overline{\Gamma}_{ab}^\text{HD}(0) = 1/3$. Throughout this paper, we normalize the HD curve as $\Gamma_{ab}^\text{HD}(0) = 1/2$, in line with the community \cite{NANOGrav:2021ini}, and measure the GW correlations relative to the HD curve. The ORF normalization is understandably an aesthetic choice, but to compare our expressions with \cite{Allen:2022dzg} we multiplied the analytical expressions by the factors of $1.5 = 0.5/0.333\cdots$, to take care of the differences in normalization. In particular, for the mean, we multiplied by $1.5$. However, for the variances we multiplied by $2 \times (1.5)^2$, instead of merely $(1.5)^2$. The extra factor of $2$ in the variances is drawn from two differences between \cite{Allen:2022dzg} and ours. First, in \cite{Allen:2022dzg}, the time averaging of the two point product $r_a r_b$ is taken, while in ours, we time average the residual $r_a$. Secondly, in \cite{Allen:2022dzg}, the various sources in the sky were also considered in the averaging. We thus confirm the agreement between the PSF result for the theoretical uncertainties, and the analytical ones in \cite{Allen:2022dzg} for the HD. We also mention that the harmonic analysis approach to the cosmic variance was briefly touched on in the latest revision of \cite{Allen:2022dzg} for the standard GR tensor case.

In the following sections, we show how the PSF naturally generalizes the computation of the variance of the ORF expected in a PTA. This is for the most general subluminal GW polarizations and arbitrary pulsar distances.

\section{Tensor polarizations}
\label{sec:tensorpols}

We start with the general tensor polarization as it departs from the HD correlation via subluminal velocities and finite pulsar distances.

Our choices of velocity are $v = 0.99, 0.50, 0.01$ which we refer to as `near luminal', `half luminal', and `nonrelativistic'. The luminal tensor case ($v = 1$) is practically indistinguishable from the HD curve for relevant angular separations in the present data set. We also consider varying distances, $fD = 100, 300, 1000$, to show how the correlation depends on this astrophysical parameter. With a reference frequency of $f = 1 \ {\rm yr}^{-1}$, utilized by the present PTAs, these distances translate to approximately $22$, $67$, and $223$ parsecs, or generally $D[{\rm pc}] = 22.3 \times fD/100$.

\begin{figure}[h!]
\center
	\subfigure[  ]{
		\includegraphics[width = 0.45 \textwidth]{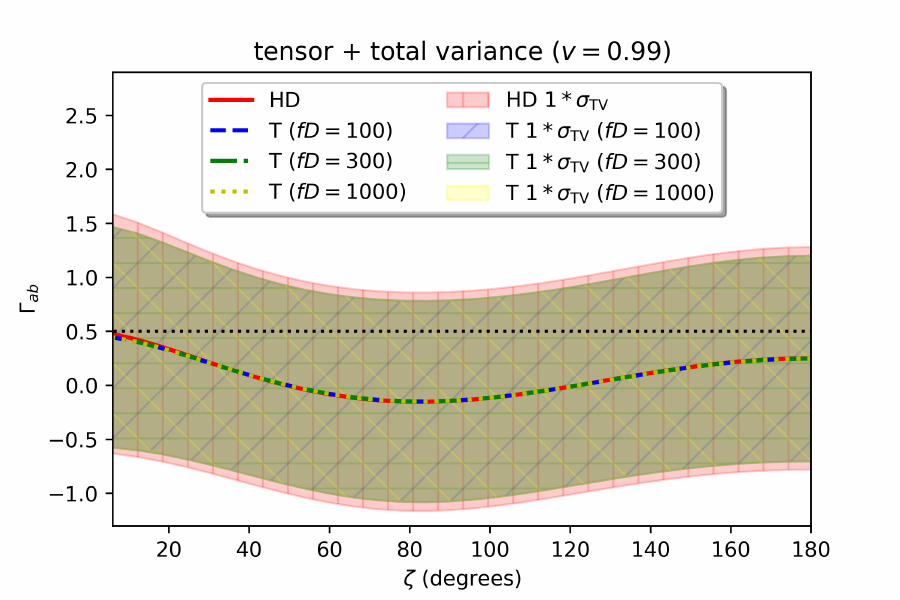}
		}
	\subfigure[  ]{
		\includegraphics[width = 0.45 \textwidth]{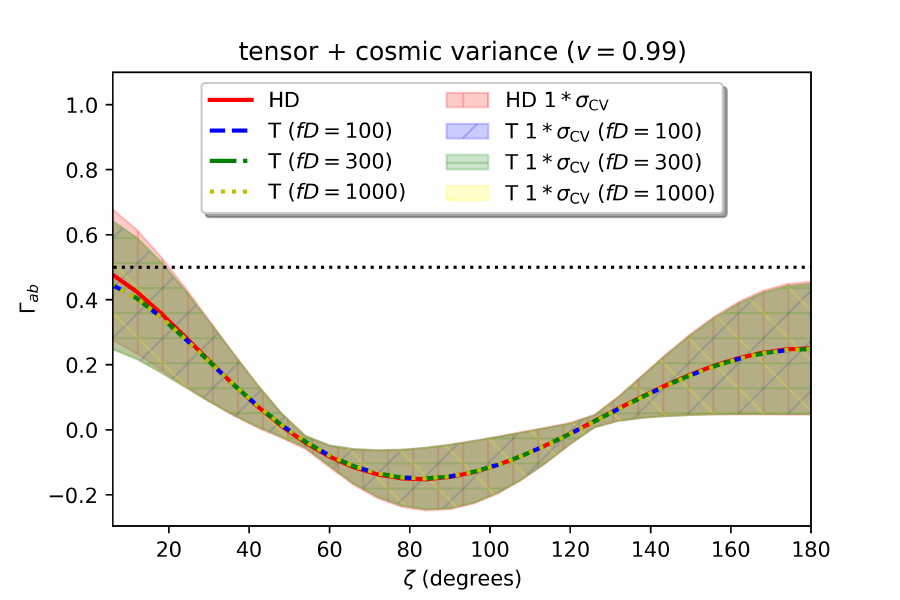}
		}
	\subfigure[  ]{
		\includegraphics[width = 0.45 \textwidth]{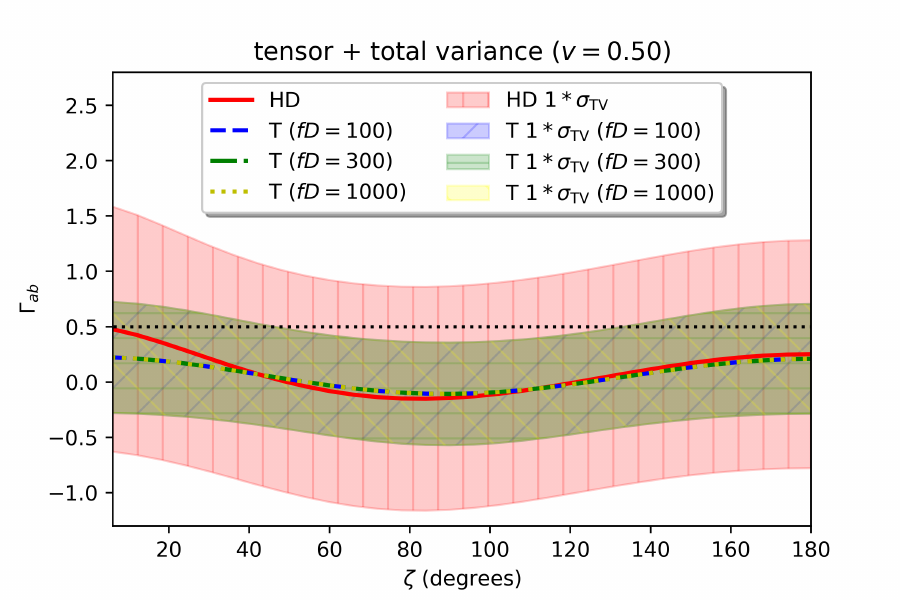}
		}
	\subfigure[  ]{
		\includegraphics[width = 0.45 \textwidth]{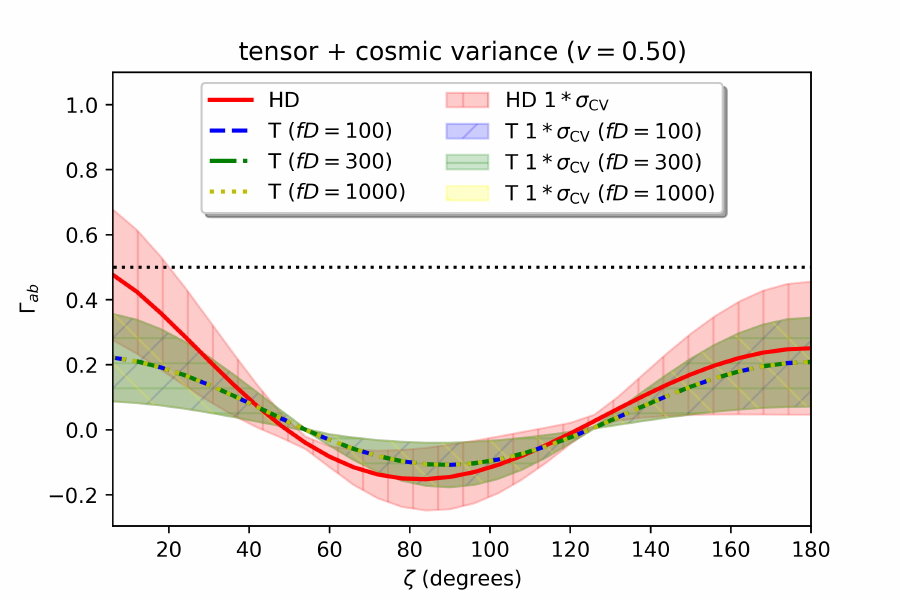}
		}
	\subfigure[  ]{
		\includegraphics[width = 0.45 \textwidth]{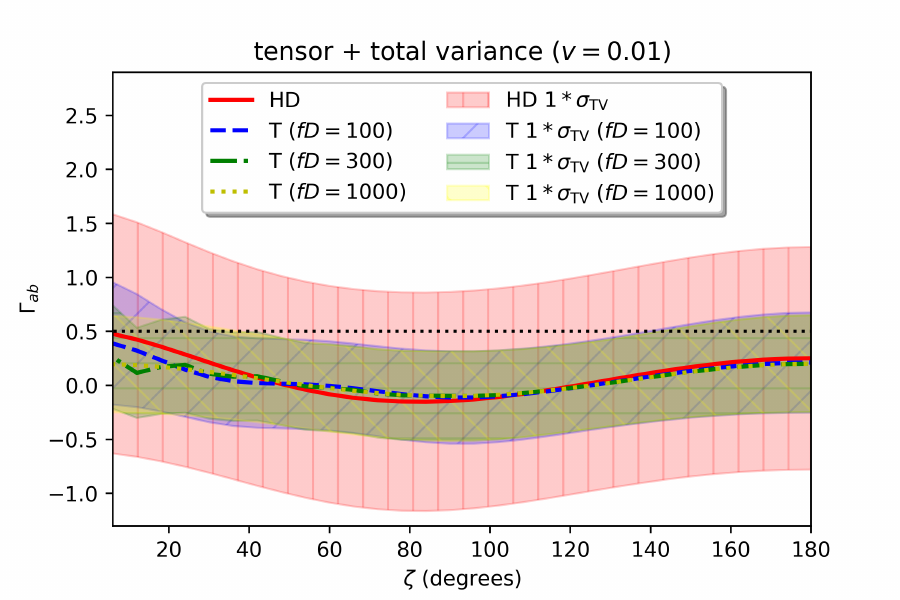}
		}
	\subfigure[  ]{
		\includegraphics[width = 0.45 \textwidth]{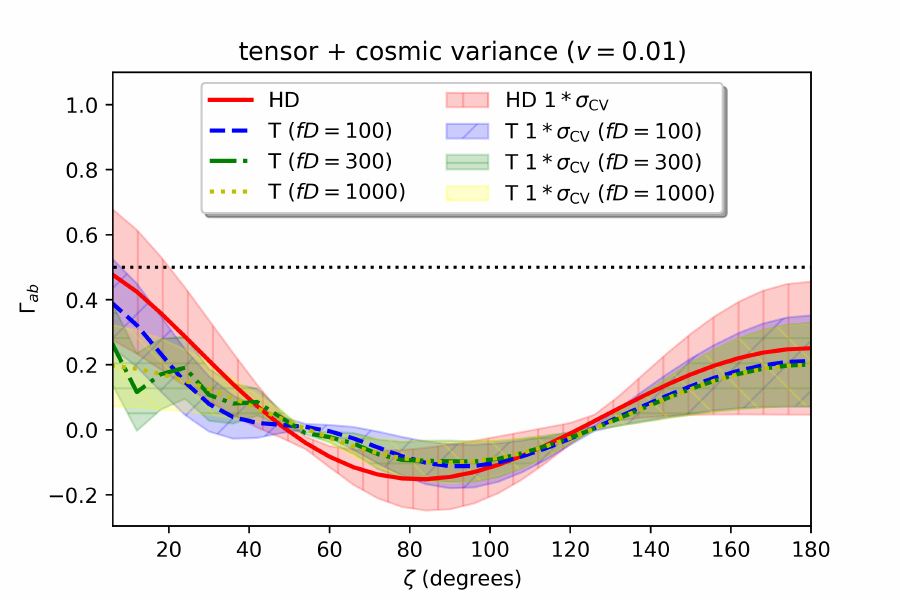}
		}
\caption{Mean and uncertainty of the ORFs for tensor modes at distances $fD = 100, 300, 1000$, and velocities $v = 0.99, 0.50, 0.01$. The first thirty multipoles ($l \leq 30$) were considered for the calculation and the autocorrelation was obtained using the real space formalism \cite{Bernardo:2022rif}. The horizontal dotted line corresponds to a `monopolar' correlation.}
\label{fig:tensorvar}
\end{figure}

Figure \ref{fig:tensorvar} shows the mean and the uncertainty separately resulting from the total and cosmic variances of the GW correlation. The HD curve and its corresponding uncertainty is shown for reference.

Since the discussion of the mean of the ORF for subluminal GW polarizations and finite distances were already given much attention in \cite{Bernardo:2022rif}, we focus our discussion on the uncertainty. First for the total variance, as with the HD, we find that this is dominated by the autocorrelation, making it visually appear like a constant away from the mean. As shown in Figure \ref{fig:tensorvar}, the total variance of a subluminal tensor GW is narrower than the corresponding variance of the HD curve, which is its luminal and infinite distance limit. The uncertainty becomes even narrower for subluminal tensor modes at half the speed of light. Still, the results also show that the distance is less of a factor compared with the velocity. This is seen in the plots, while the velocity is fixed and the distance varies. Concretely the mean and uncertainty are indistinguishable for the near luminal and half luminal tensor modes with varying pulsar distances. The situation changes for the nonrelativistic modes, as with the mean \cite{Bernardo:2022rif}. In this speed limit, the power spectrum profile becomes dominated not just by the dipole and the quadrupole, but also by the other low multipoles, hence resulting in a nontrivial ORF at large scales. In other words, these significant low multipoles at nonrelativistic speeds exhibit themselves as an oscillatory feature in the spatial correlation at large scales, and carry a stronger distance dependence that allows the ORFs to be distinguishable for different distances \cite{Bernardo:2022rif}. This similarly influences the total variance, for which we now find the curves to be distinguishable with the distance.

On to the cosmic variance, we find some similar trends. For one, we see the uncertainty of the general subluminal tensor to be narrower than that of the HD curve. This is shown in the near luminal and half luminal tensor modes in Figure \ref{fig:tensorvar}. The pulsar distances are also not so much a factor as with the velocity, except at nonrelativistic speeds where the low multipoles other than the quadrupole contribute significantly to the total power. On the other hand, the minima of the cosmic variance are roughly at the same places as with the HD, regardless of the velocity and the distance. This information may be utilized to find departures of tensor anchored SGWB from observations, as the predictions for pulsar pairs with such angular separations become more certain.

We note that the low angle oscillations at nonrelativistic speeds is sourced by an enhanced higher order multipolar power spectrum. This invites the question: ``Would the tensor curves look different at small angles compared to large angles?'' The general answer is yes. The caveat to this is that it requires an incredible observational precision to resolve this since the difference appears at small angles. For example, for thirty, fifty, or the first hundred multipoles, the nonrelativistic curves look exactly like Figure \ref{fig:tensorvar}(e-f) in the region $\zeta \leq 6^\circ = 180^\circ/30$, including the uncertainties. Thirty multipoles is a conservative choice for the present data set. By and large, seeing the oscillations or other features induced by higher multipoles, $l \sim l_{\rm max}$, requires resolving pulsars of separations $\zeta \leq 180^\circ/l_{\rm max}$. We look forward to future PTA missions to look for these signatures.

The present PTA data have quite large uncertainties owing to monopolar spatial correlations being quite statistically significant. This is understandably because of limitations in the optimal statistic analysis, and the spatially correlated monopole not being independent of the spatially uncorrelated common spectrum process \cite{NANOGrav:2020bcs}. The observation of about a thousand millisecond pulsars, as targeted by future PTA missions, is expected to narrow down the uncertainty and most importantly host data points at low angles where the tensor ORFs are distinguishable with each other and the HD curve.

We end this section by mentioning the possibility of subluminal GWs as raised by dark energy \cite{deRham:2018red}. Notwithstanding the little wiggle room for error on the GW speed, due to the astounding observation of GWs and gamma ray bursts from a neutron star binary \cite{LIGOScientific:2017vwq}, this measurement takes place near effective field theory cutoff of dark energy. It may well be the case that GWs propagate at a different speed than light in vacuum, but only by chance go as fast as light in the frequency band of ground based GW observatories. Whether such a dispersive nature can be observed in the millihertz GW band is up to the space based detectors, while the nanohertz band is up to PTAs. Of course, before any science inference, the modelling must be assembled.

\section{Vector polarizations}
\label{sec:vectorpols}

We now take a look at vector polarizations. We consider the same choices of the velocity $v = 0.99, 0.50, 0.01$ and distances $fD = 100, 300, 1000$ as with the tensor modes.

\begin{figure}[h!]
\center
	\subfigure[  ]{
		\includegraphics[width = 0.45 \textwidth]{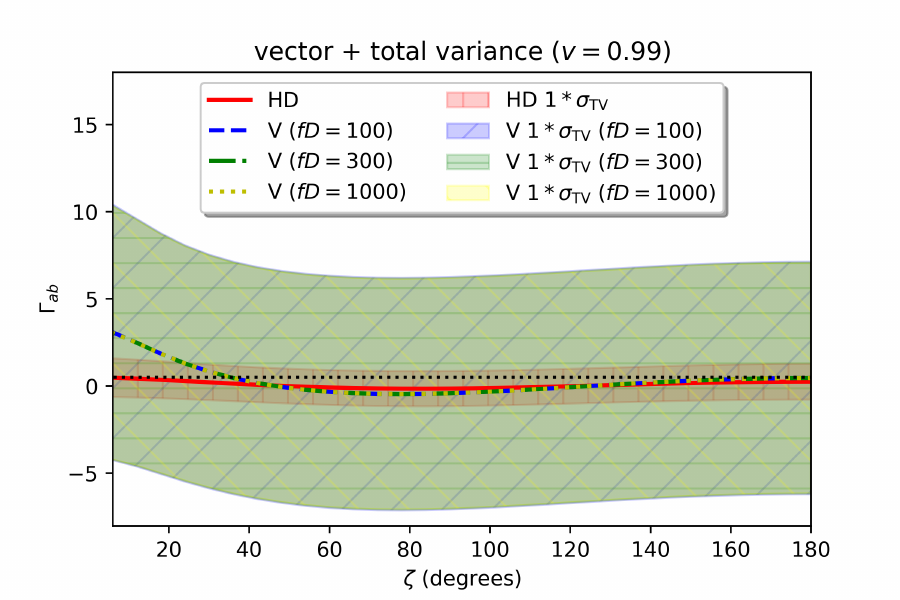}
		}
	\subfigure[  ]{
		\includegraphics[width = 0.45 \textwidth]{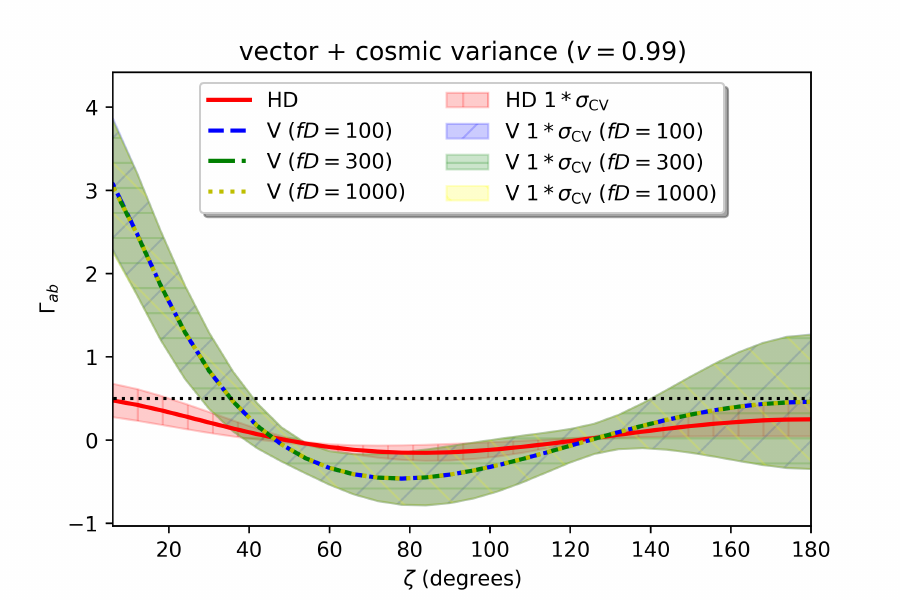}
		}
	\subfigure[  ]{
		\includegraphics[width = 0.45 \textwidth]{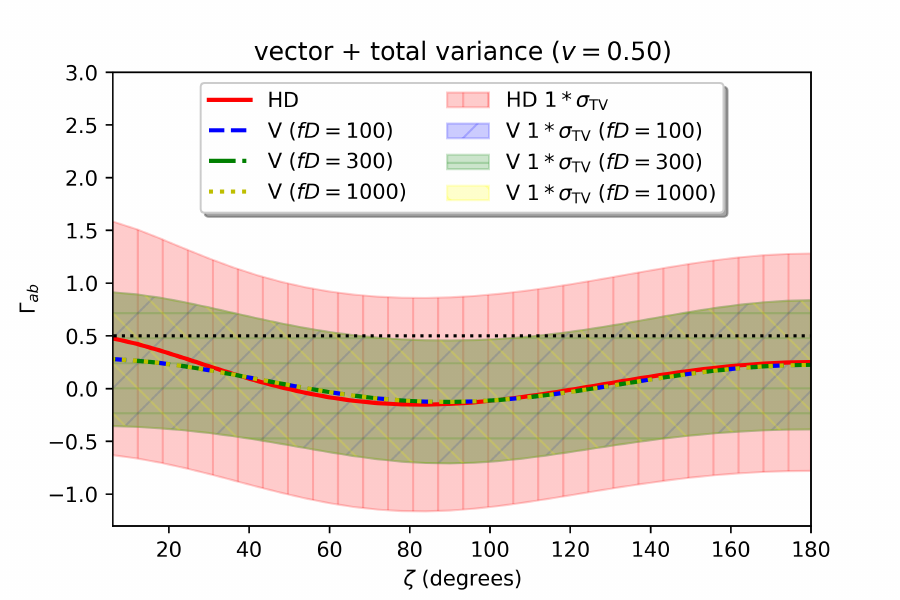}
		}
	\subfigure[  ]{
		\includegraphics[width = 0.45 \textwidth]{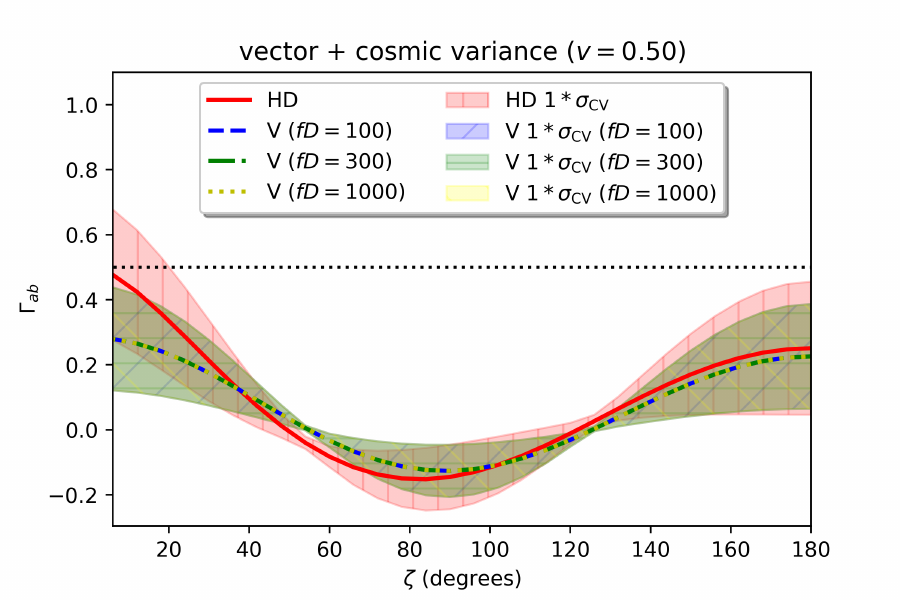}
		}
	\subfigure[  ]{
		\includegraphics[width = 0.45 \textwidth]{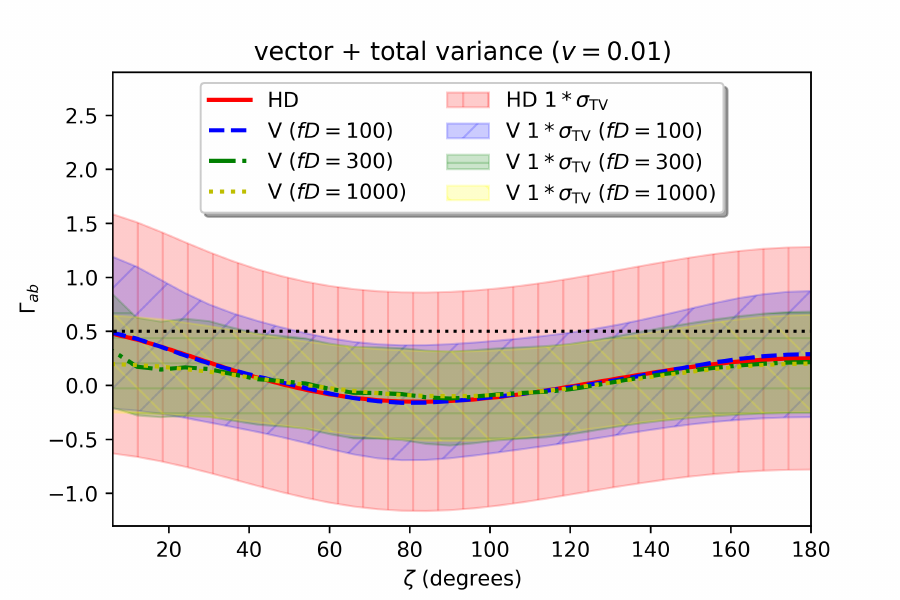}
		}
	\subfigure[  ]{
		\includegraphics[width = 0.45 \textwidth]{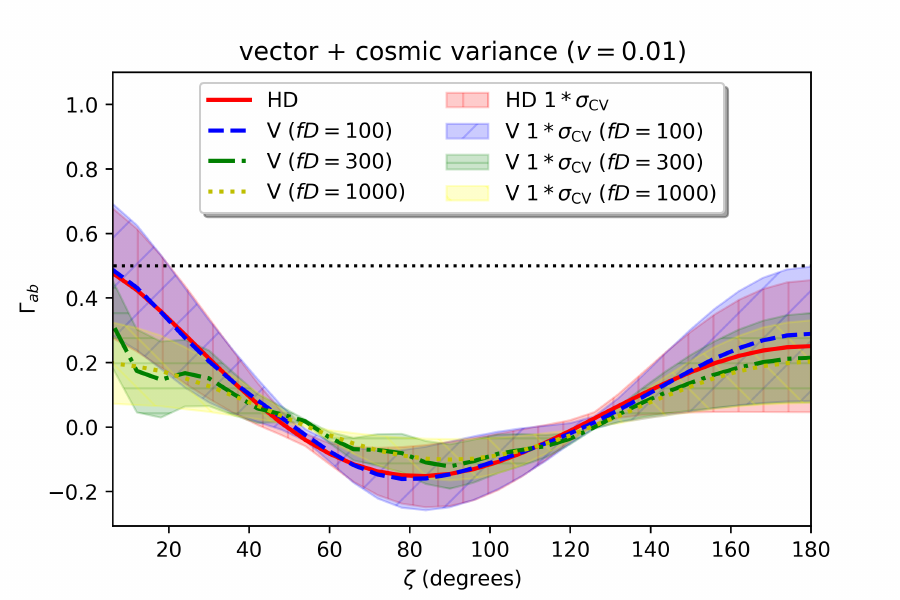}
		}
\caption{Mean and uncertainty of the ORFs for vector modes at distances $fD = 100, 300, 1000$, and velocities $v = 0.99, 0.50, 0.01$. The first thirty multipoles ($l \leq 30$) were considered for the calculation and the autocorrelation was obtained using the real space formalism \cite{Bernardo:2022rif}. The horizontal dotted line corresponds to a `monopolar' correlation.}
\label{fig:vectorvar}
\end{figure}

We start by reminding that the vector modes' ORF diverge in the luminal and infinite distance limit \cite{NANOGrav:2021ini}, and so are not compatible with observations. We shall find the residue of this divergence in the near luminal case. Figure \ref{fig:vectorvar} shows the total and cosmic variance uncertainties of the vector ORFs and the corresponding HD correlation.

Focusing on the near luminal vector, we find rather large total variances owing to the autocorrelation reflecting residues of the divergence in the luminal and infinite pulsar distance limit. As mentioned, this deprives it from any predictive power and is also incompatible with observations. Nonetheless at smaller speeds, the ORF predictions  for the vector polarizations become more reliable. This time, at half luminal speed for instance, the total variances can be found inside the HD one. The pulsar distance also is clearly not so much of a factor, as the different ORFs for various distances visually overlap. This goes on for the near luminal down to half luminal cases. At nonrelativistic speeds, on the other hand, the vector ORFs, and the corresponding total variances, become distinguishable at large angles, similar with the tensor counterpart. This is due to the power spectrum being significantly populated by low multipoles aside from the dipole and the quadrupole \cite{Bernardo:2022rif}. Observational data at small pulsar pair angular separations should then be able to distinguish even between nonrelativistic vector cases.

Taking the discussion to the cosmic variance, which again results from the full sky averaging across pulsar pairs of fixed angular separation, we see a significant departure of the near luminal vector ORFs from the HD curve by more than a few sigmas, particularly at angles below $\zeta \sim 30^\circ$. This disfavors the near luminal vector modes, as data points as far below as $\zeta \sim 14^\circ$ remain to be consistent with the HD \cite{NANOGrav:2020bcs, NANOGrav:2021ini}. We emphasize this is true as well regardless of the choice of the pulsar distance, as shown in Figure \ref{fig:vectorvar}.

At half luminal speed, we find that the ORF, taking into account the cosmic variance, becomes more consistent with the HD, except at low angles where some deviation appears. This may be distinguishable in future observations, with emphasis at low angular separations, but this data set has yet to arrive. The different choices of the pulsar distances lead to visually overlapping ORFs, showing that this parameter is not much of a factor at this velocity limit. This changes at nonrelativistic speeds. As the means become distinguishable in this speed limit, so does the uncertainties. More interestingly, we find the degeneracy in the spatial correlation sourced by the HD, that is transverse traceless tensor modes, and nonrelativistic vector polarization to extend to the cosmic variance. Of course, this appears at a very particular range of the distance, $fD \sim 100$, and so is likely ruled out when the distance data are taken into account, but this is nonetheless an interesting point to bring up.

Before we move on to scalars, we mention that vector gravitational degrees of freedom, that may source vector metric polarizations in the SGWB, are diluted by the cosmic expansion. Early sources that left their marks on the nanohertz GW sky may nonetheless persist \cite{Liang:2021bct, Tachinami:2021jnf}.

\section{Scalar polarizations}
\label{sec:scalarpols}

We tackle scalar GW polarizations in the SGWB. In contrast with the tensor and vector cases discussed previously, we separate the discussion of the scalar between two kinds: `scalar transverse' (ST) and `scalar longitudinal' (SL), where the ST distorts the space, hence test masses, perpendicular to the direction of propagation of a GW, while the SL moves masses along the GW path.

\begin{figure}[h!]
\center
	\subfigure[  ]{
		\includegraphics[width = 0.45 \textwidth]{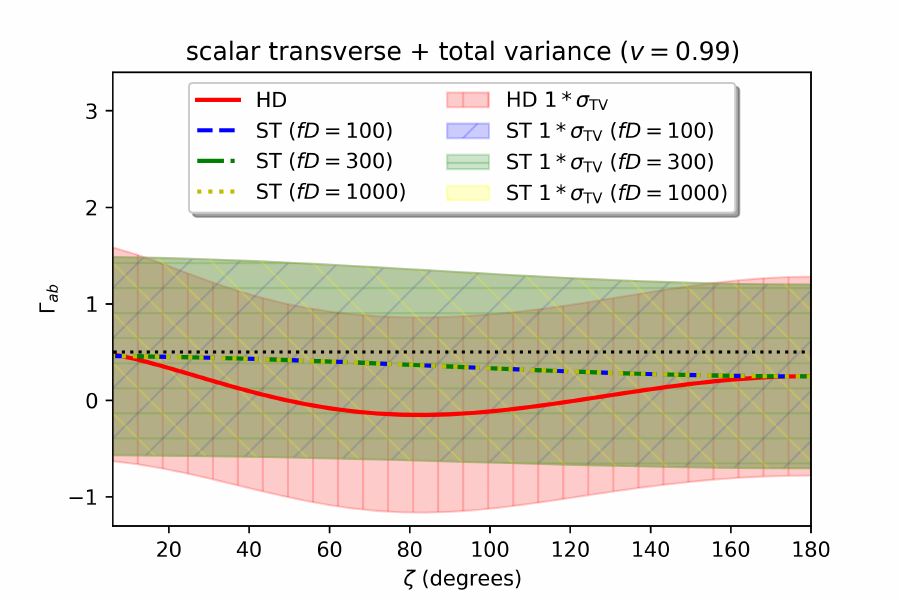}
		}
	\subfigure[  ]{
		\includegraphics[width = 0.45 \textwidth]{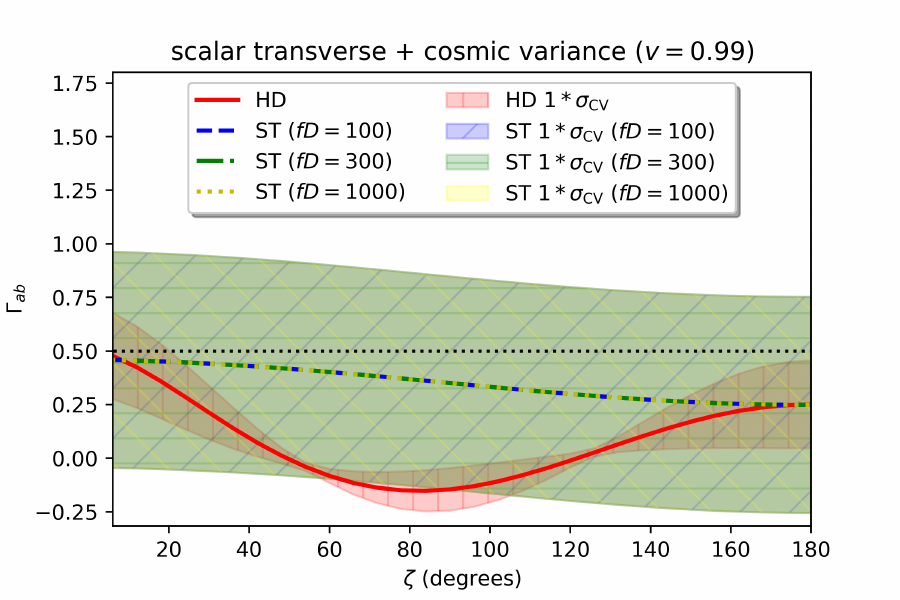}
		}
	\subfigure[  ]{
		\includegraphics[width = 0.45 \textwidth]{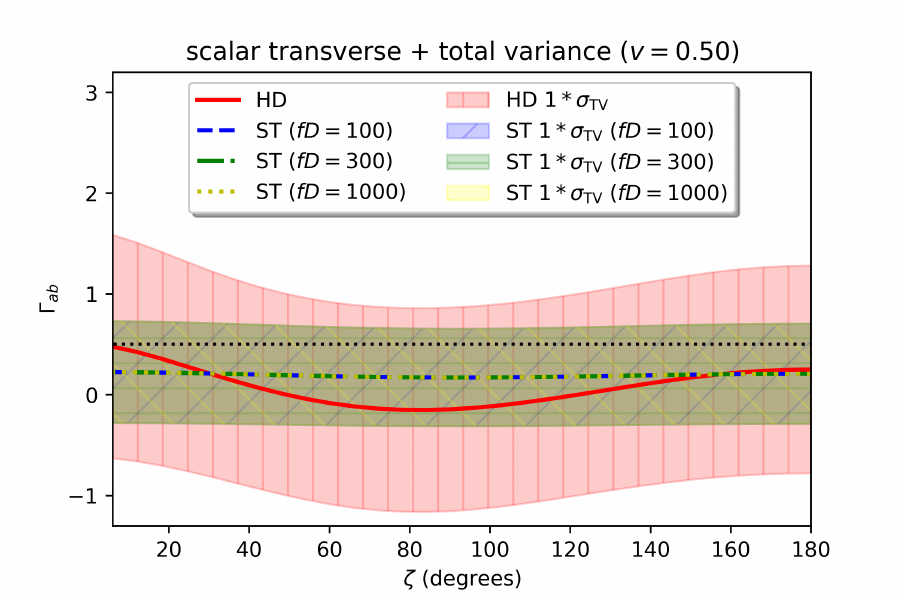}
		}
	\subfigure[  ]{
		\includegraphics[width = 0.45 \textwidth]{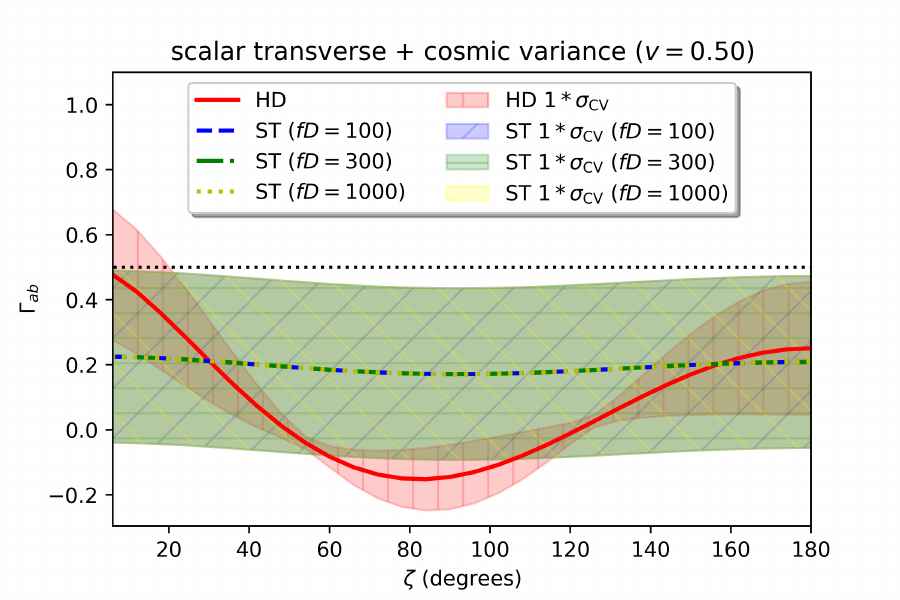}
		}
	\subfigure[  ]{
		\includegraphics[width = 0.45 \textwidth]{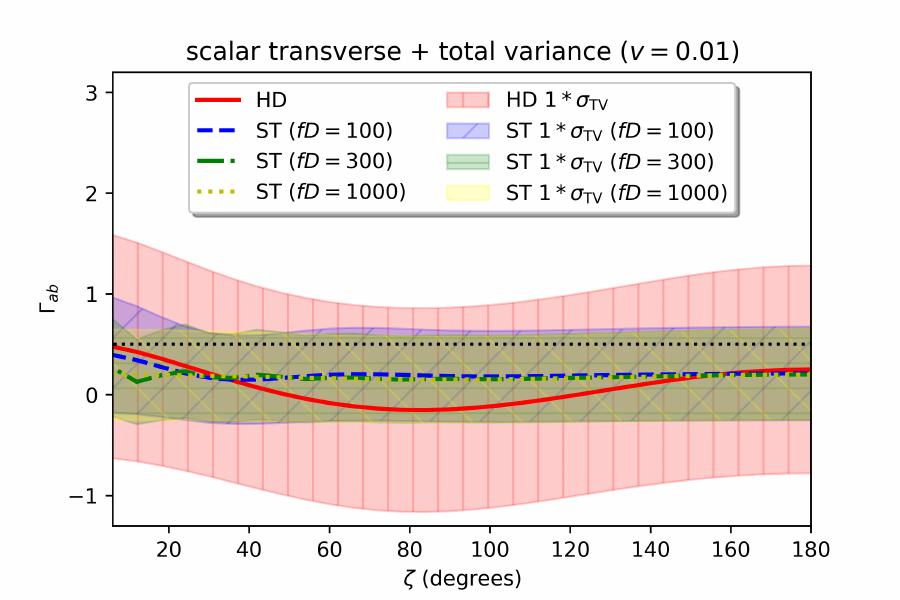}
		}
	\subfigure[  ]{
		\includegraphics[width = 0.45 \textwidth]{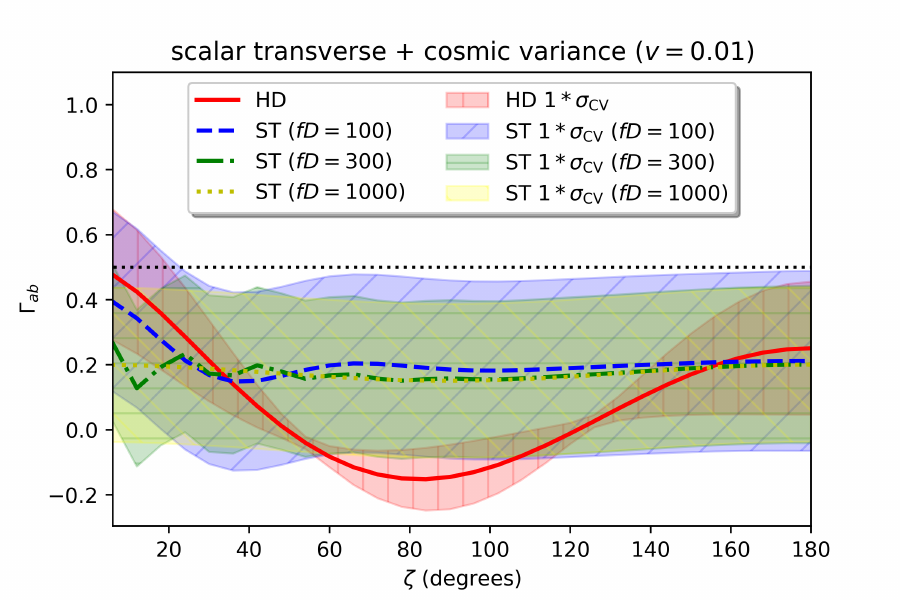}
		}
\caption{Mean and uncertainty of the ORFs for scalar transverse modes at distances $fD = 100, 300, 1000$, and velocities $v = 0.99, 0.50, 0.01$. The first thirty multipoles ($l \leq 30$) were considered for the calculation and the autocorrelation was obtained using the real space formalism \cite{Bernardo:2022rif}. The horizontal dotted line corresponds to a `monopolar' correlation.}
\label{fig:STvar}
\end{figure}

We start with the ST. Figure \ref{fig:STvar} shows the ORFs and the corresponding uncertainties for the velocities $v = 0.99, 0.50, 0.01$ and distances $fD = 100, 300, 1000$.

We remind here that the scalar power spectrum is dominated by the monopole and the dipole. This reflects in the mean ORFs in Figure \ref{fig:STvar} which are dipole-like at near luminal speed while becoming more monopole-like, that is flatter, at subluminal speeds such as at the half luminal case. At nonrelativistic speeds, the power spectrum becomes contaminated by the low order multipoles beyond the monopole and the dipole. Nonetheless, the monopolar and dipolar powers dominate even when the other multipoles contribute significantly.

For the total variance of the ST, we find that the one by the near luminal ST is only slightly narrower compared with the HD one. This is due to the ST's autocorrelation coinciding with the same final integral expression as the one by the transverse tensor modes \cite{Bernardo:2022rif}. As the autocorrelation dominates the total variance, this explains why the ST uncertainty is a bit narrower than the HD. A main difference however is the shape of the means of the ORFs of the ST. This makes them visually distinguishable with the HD, even though these are not distinguishable within themselves with varying pulsar distances. Now, as the autocorrelation decreases at subluminal velocities, we find that the uncertainty also becomes even more narrower. This is shown at half luminal in Figure \ref{fig:STvar} where since the monopolar power dominates, the ORF looks like a mere flat line with flat uncertainty. The distance dependence remains mild at this speed, echoing what we found with the tensor and vector polarizations. This changes again at nonrelativistic speeds where we find nontrivial spatial correlation and uncertainty due to the power spectrum being contaminated by low multipoles beside the monopole and the dipole. In this case, the ORFs for different pulsar distances become somewhat distinguishable especially at low angles. However, their means are all within one sigmas of each other, showing that the distance dependence of the GW correlation is really just quite small when taking into account the uncertainties. 

We move the discussion to the cosmic variance of the ST. This is shown in Figure \ref{fig:STvar} for the same velocities and pulsar distances. What is most interesting here is that now there are no more apparent minima in the cosmic variance, as the HD, tensor, and vector polarizations present. This can be traced to the power spectrum being dominated by the monopole and the dipole, or rather that this time the quadrupolar component is suppressed. Interestingly, the cosmic variance of the ST is also generally larger than the HD one, except at nonrelativistic speeds which we shall go back to in a moment. Meanwhile, we see that the pulsar distances can be regarded to be not much of a factor when the uncertainties are taken into account. This is even more true of the luminal and half luminal ST polarization where the cosmic variance is quite large. At nonrelativistic speeds, the ORF behavior becomes quite interesting, once more as the pulsar distance factors in the low angle correlation. The means are inside the others cosmic variance regardless and so can be taken to imply that the distance dependence is only mild. We find that the cosmic variance of the HD is generally narrower than that of the ST polarization.

\begin{figure}[h!]
\center
	\subfigure[  ]{
		\includegraphics[width = 0.45 \textwidth]{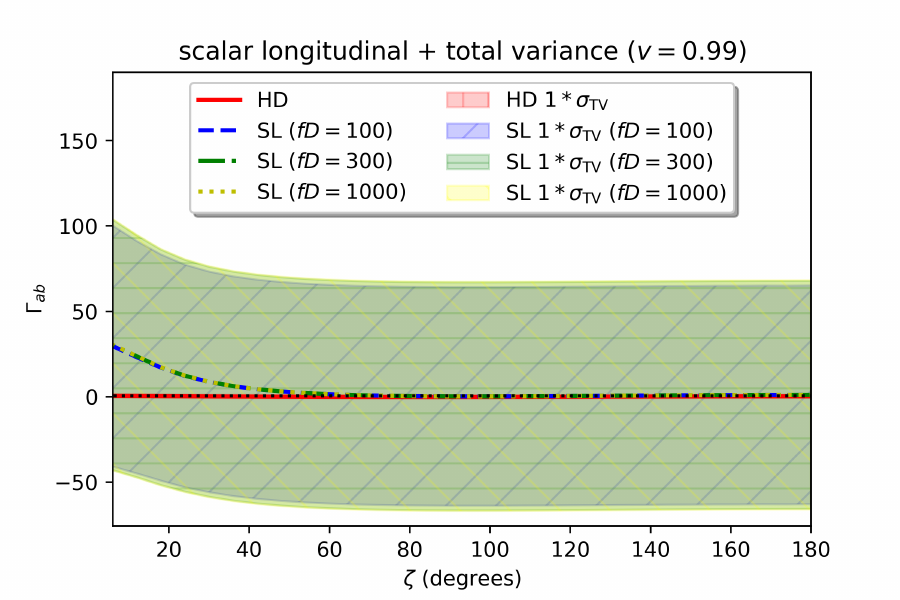}
		}
	\subfigure[  ]{
		\includegraphics[width = 0.45 \textwidth]{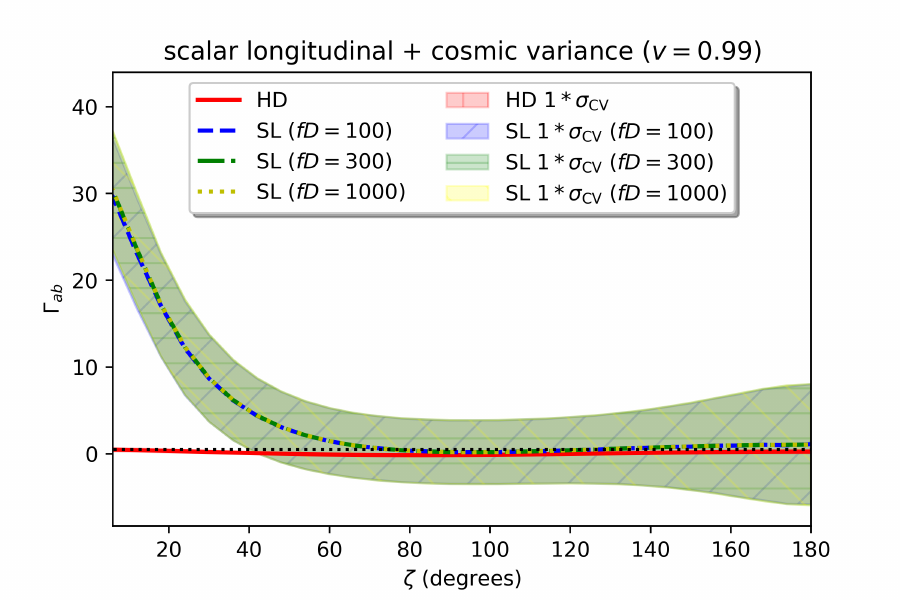}
		}
	\subfigure[  ]{
		\includegraphics[width = 0.45 \textwidth]{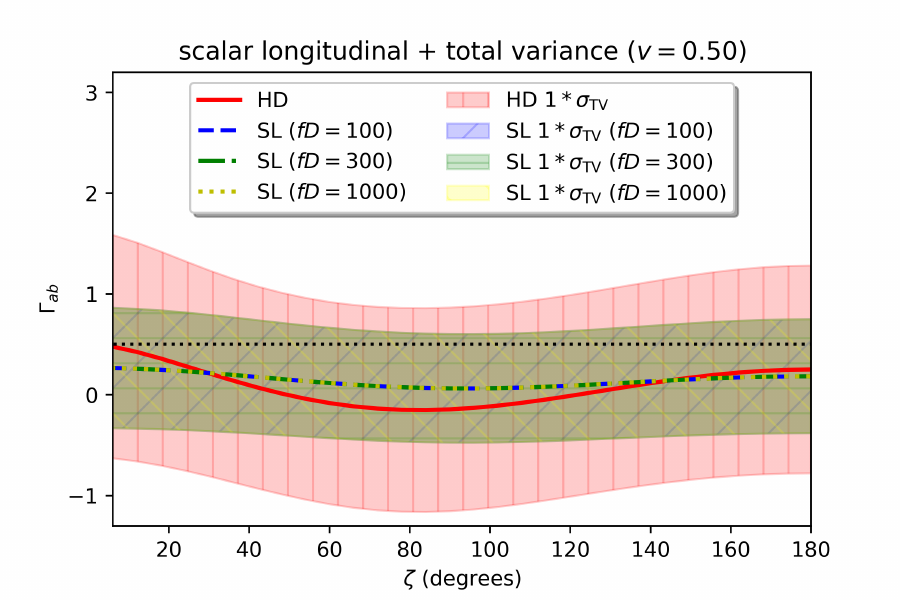}
		}
	\subfigure[  ]{
		\includegraphics[width = 0.45 \textwidth]{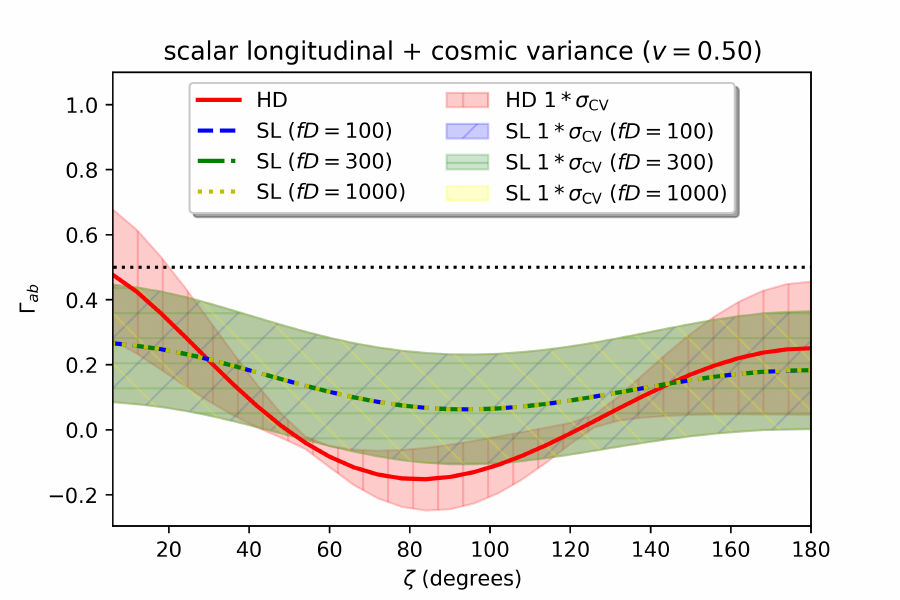}
		}
	\subfigure[  ]{
		\includegraphics[width = 0.45 \textwidth]{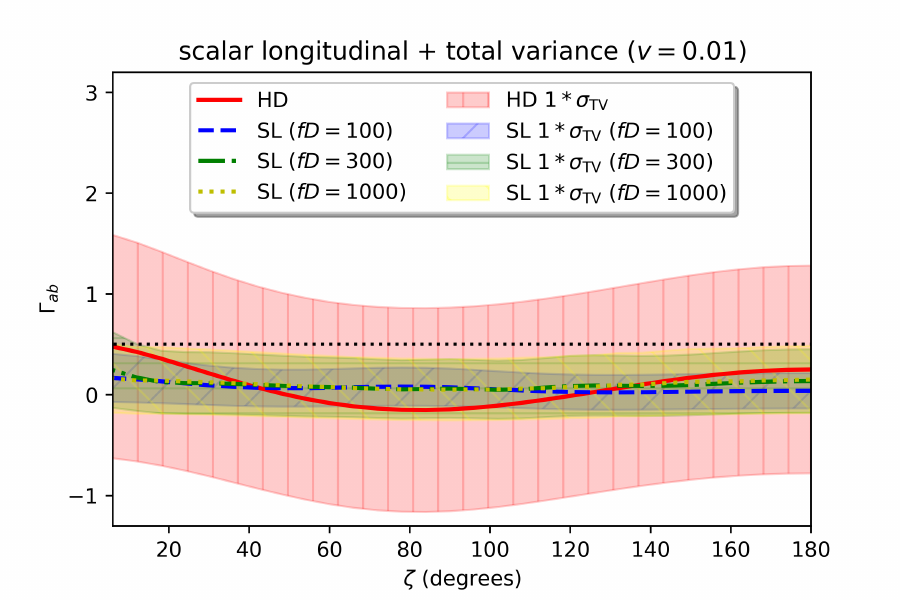}
		}
	\subfigure[  ]{
		\includegraphics[width = 0.45 \textwidth]{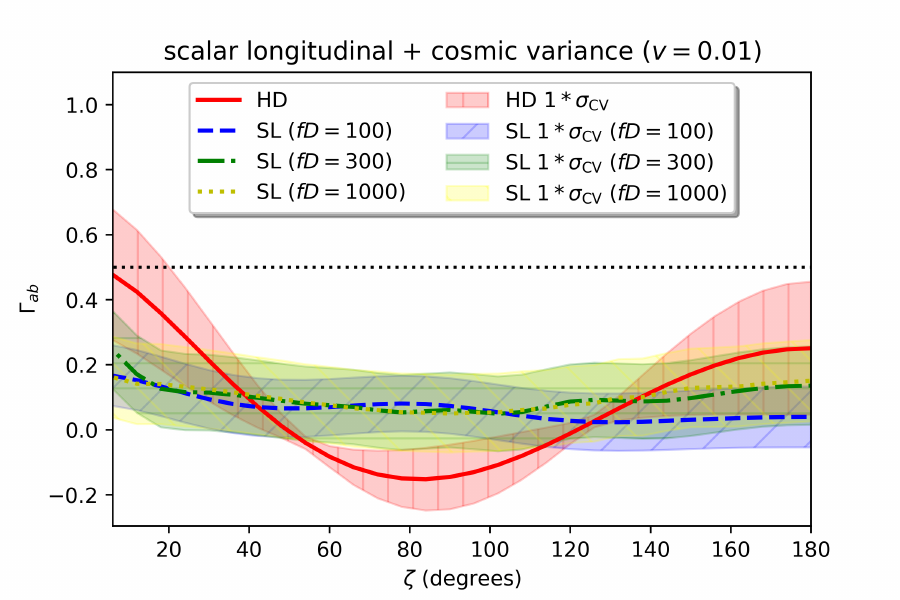}
		}
\caption{Mean and uncertainty of the ORFs for scalar longitudinal modes at distances $fD = 100, 300, 1000$, and velocities $v = 0.99, 0.50, 0.01$. The first thirty multipoles ($l \leq 30$) were considered for the calculation and the autocorrelation was obtained using the real space formalism \cite{Bernardo:2022rif}. The horizontal dotted line corresponds to a `monopolar' correlation.}
\label{fig:SLvar}
\end{figure}

We last move to the SL polarization. First, we mention that like the vector case, the SL modes' ORF diverges, becomes undefined, in the luminal and infinite distance limit. This will leave a residue in the near luminal case as we are about to see. Figure \ref{fig:SLvar} shows the ORFs and the corresponding uncertainties sourced by the SL polarization for the various velocities and distances as used previously with the ST.

For the total variance, we find, as a residue of the divergence in the luminal and infinite distance limit, that the near luminal SL's spatial correlation just loses predictability, or in other words, comes with very large uncertainty. This rules it out with respect to the observation \cite{NANOGrav:2021ini}. Nonetheless, the SL modes' ORFs become meaningful at subluminal speeds. At half luminal for instance, its ORF shapes quite like a dipole but with some added quadrupolar component. The SL's uncertainty is also narrower than the HD, in stark contrast with the ST case. At near luminal and half luminal speeds, the ORFs corresponding to different pulsar distances can be seen to be merely overlapping. A deviation to this can be found at nonrelativistic speeds, where in addition to the uncertainty becoming even narrower, due to the nontrivial power spectrum profile in this limit, the different ORFs corresponding to different distances become visually distinguishable. However, now taking the uncertainty in consideration, we see that each finite pulsar distance predictions are within the others. This means that the GW correlation just mildly depends on the pulsar distances. As we are observing this for all metric polarizations, this seems to be quite a general statement.

Lastly, for the SL modes cosmic variances, we see in Figure \ref{fig:SLvar} that the near luminal SL mode can be regarded as disfavored as it departs too significantly from the HD \cite{NANOGrav:2021ini}. The SL's prediction is better away from the light cone. As can be seen, at half luminal speed, the SL's cosmic variance may be seen to be even as wide as that of the HD at small angles, except it does not reflect the minima in the cosmic variance that tensor and vector polarizations express. Once more, the predictions are overlapping for different choices of the pulsar distances, which only show how little the choice of this parameter matters at these speeds. The pulsar distance dependence instead kicks in at nonrelativistic speeds. Down this speed limit, visual differences can be seen in both the mean and uncertainty of the ORFs of various distances. This is once again a reflection of the nontrivial power spectrum profile of the SL in this speed limit. The uncertainties of each ORF however manage to include the means of the other predictions at different distances. This suggests that the pulsar distances are not so much of a factor in the GW correlation, as we found for all the other metric polarizations.

To end the section, we comment that the scalar GW polarizations are triggered by scalar gravitational degrees of freedom, for example, in Galileon scalar field theories. The two modes, ST and SL, generally come as a mixture in the SGWB, where the amount of mixing is determined by the mass of the propagating degree of freedom, which is directly related to the velocity. To be concrete, in both $f(R)$ and the Galileon, it turns out that the mixture is specified by the constant $r = (1 - v^2)/\sqrt{2}$ such that the GW comes in the combination $h_{ab}^\text{ST} + r h_{ab}^\text{SL}$ \cite{Qin:2020hfy, Bernardo:2022vlj}. Restricting the scalar field on the light cone therefore kills off the SL mode, which diverges in that speed limit, but in general both modes enter with certain amounts. The Galileon was constrained in \cite{Bernardo:2022vlj}, demonstrating the numerical efficiency of the power spectrum method as well as the potential of SGWB spatial correlations in a PTA to constrain modified gravity degrees of freedom.

\section{Discussion}
\label{sec:discussion}

We have advocated the power spectrum formalism as a numerically efficient tool in studying the stochastic gravitational wave background correlations in a pulsar timing array. In this paper, we have further shown that it can be used to straightforwardly obtain the theoretical uncertainties that may show up in pulsar timing array observations due to the way pulsars are scattered across the sky. This can be done easily for any gravitational wave polarization, propagating on or off the light cone, and with arbitrary pulsar distances \cite{Bernardo:2022rif}.

We have found that the pulsar distances from the observer do not play too much of a role when the uncertainty is taken into account. This however does not mean that the pulsar distance should be taken to infinity, as such the luminal vector and scalar modes make divergent predictions, and that the nonrelativistic scalar modes get contaminated by unrealistically large monopolar and dipolar components \cite{Bernardo:2022rif}. Keeping astrophysical finite pulsar distances in the simulations has the advantage of making \textit{all} the polarizations well defined, which is suitable for an efficient numerical implementation that can be used for data analysis of pulsar timing array observations.

As did the cosmic microwave background, we anticipate pulsar timing array endeavor to also grow into a more mature science, able to make precise inferences about the physical nature of the sources of the stochastic gravitational wave background. The observation of about a thousand millisecond pulsars, as targeted in future pulsar timing array missions \cite{Smits:2008cf, Lazio:2013mea, Janssen:2014dka, Miao:2021awa, Joshi:2022ohi, Bustamante-Rosell:2021daj}, should be able to improve the measurement uncertainties. Observations of pulsar pairs of subdegree separations in particular is something to look forward to as significant departures from the conservative Hellings-Downs correlation can be expected in this regime (forecasted in Figure \ref{fig:hdlowangle}, Section \ref{sec:hdmeanvar}). It is in this low angle region that the overlap reduction functions sourced by alternative gravity degrees of freedom become distinguishable, not just by their velocities but also by the astrophysical distance of the pulsars (Sections \ref{sec:tensorpols}, \ref{sec:vectorpols}, and \ref{sec:scalarpols}). If any, we can attest the science we have yet to learn about nanohertz GWs to be an exciting one, full of surprises about the early cosmic history. The power spectrum formalism paves the road to do so by means of spatial correlation physical predictions, with a mean and an uncertainty.

We setup some future directions, beginning with the timeliest one, writing a computational package that provides the overlap reduction function given a set of pulsar distances, and gravitational wave polarizations and propagation speeds. This prepares the implementation of subluminal stochastic gravitational wave background in the present statistical framework in pulsar timing array data analysis. Along this line, it will be interesting to see whether evidence could be obtained for subluminal modes, hinting at the possible dispersive nature of gravitational waves. On the theoretical side, this calls for theorists to draw up constraints on the parameter space of alternative gravity theories with pulsar timing array observations. It also remains to setup a general formalism for possible scalar and vector mode induced anisotropies in the stochastic gravitational wave background, much like \cite{Liu:2022skj} for luminal tensor polarizations. While this is way ahead of its time, surely measuring the anisotropies in the stochastic gravitational wave background is an exciting scientific prospect.

\section*{Acknowledgements}
The authors thank Bruce Allen for getting in touch about the harmonic analysis approach to the cosmic variance for the standard GR case. This work was supported in part by the Ministry of Science and Technology (MOST) of Taiwan, Republic of China, under Grant No. MOST 111-2112-M-001-065.

\appendix

\section{Technical derivation of the pulsar variances}
\label{sec:technical_derivations}

We present the more detailed steps in the derivation of the total and cosmic variances in Section \ref{sec:pta_variances_psmeth}

\subsection{Total variance}
\label{subsec:totalvar_appendix}

To obtain the variance (\ref{eq:variance}), we calculate the second moment of the correlation operator.

We start by writing down the explicit form
\begin{equation}
    \left( {\bm \beta}_a^\dagger \bm{\beta}_b \right)^2 = \sum_{l_1 m_1} {\bm b}^\dagger_{l_1 m_1} Y^*_{l_1 m_1}\left(\hat{e}_a\right) \sum_{l_2 m_2} {\bm b}_{l_2 m_2} Y_{l_2 m_2}\left(\hat{e}_b\right) \sum_{l_3 m_3} {\bm b}^\dagger_{l_3 m_3} Y^*_{l_3 m_3}\left(\hat{e}_a\right) \sum_{l_4 m_4} {\bm b}_{l_4 m_4} Y_{l_4 m_4}\left(\hat{e}_b\right) \,.
\end{equation}
Taking the ensemble average of this leads to
\begin{equation}
    \langle \left( {\bm \beta}_a^\dagger \bm{\beta}_b \right)^2 \rangle = \sum_{l_1 m_1} \sum_{l_2 m_2} \sum_{l_3 m_3} \sum_{l_4 m_4} \langle {\bm b}^\dagger_{l_1 m_1} {\bm b}_{l_2 m_2} {\bm b}^\dagger_{l_3 m_3} {\bm b}_{l_4 m_4} \rangle Y_{l_1 m_1}^* \left(\hat{e}_a\right) Y_{l_2 m_2} \left(\hat{e}_b\right) Y_{l_3 m_3}^* \left(\hat{e}_a\right) Y_{l_4 m_4} \left(\hat{e}_b\right) \,.
\end{equation}
We press on by considering that the fields involved are statistically Gaussian. This way, the four point function Wick rotates, or factorizes, into a product of two point functions as
\begin{equation}
\label{eq:wickrotation}
\begin{split}
    \langle {\bm b}^\dagger_{l_1 m_1} {\bm b}_{l_2 m_2} {\bm b}^\dagger_{l_3 m_3} {\bm b}_{l_4 m_4} \rangle = \ &  C_{l_1} C_{l_3} \delta_{l_1 l_2} \delta_{m_1 m_2} \delta_{l_3 l_4} \delta_{m_3 m_4} + C_{l_1} C_{l_2} \delta_{l_1 l_4} \delta_{m_1 m_4} \delta_{l_2 l_3} \delta_{m_2 m_3} \\
    & + (-1)^{m_1} (-1)^{m_2} C_{l_1} C_{l_2} \delta_{l_1 l_3} \delta_{m_1 -m_3} \delta_{l_2 l_4} \delta_{m_2 -m_4} \,.
\end{split}
\end{equation}
Taking this into account, after performing the sum over the Kronecker deltas, the second moment of $\bm{\gamma}_{ab}$ becomes
\begin{equation}
\label{eq:secondmomenthd_almost}
\begin{split}
    \langle \left( {\bm \beta}_a^\dagger \bm{\beta}_b \right)^2 \rangle = \ & \sum_{l_1 m_1} \sum_{l_3 m_3} C_{l_1} C_{l_3} Y_{l_1 m_1}^*\left( \hat{e}_a \right) Y_{l_1 m_1}\left( \hat{e}_b \right) Y_{l_3 m_3}^*\left( \hat{e}_a \right) Y_{l_3 m_3}\left( \hat{e}_b \right) \\
    & \phantom{gg} + \sum_{l_1 m_1} \sum_{l_2 m_2} C_{l_1} C_{l_2} Y_{l_1 m_1}^*\left( \hat{e}_a \right) Y_{l_1 m_1}\left( \hat{e}_b \right) Y_{l_2 m_2}\left( \hat{e}_b \right) Y_{l_2 m_2}^*\left( \hat{e}_a \right) \\
    & \phantom{gg} + \sum_{l_1 m_1} \sum_{l_2 m_2} (-1)^{m_1} (-1)^{m_2} C_{l_1} C_{l_2} Y_{l_1 m_1}^*\left( \hat{e}_a \right) Y^*_{l_1 -m_1}\left( \hat{e}_a \right) Y_{l_2 m_2}\left( \hat{e}_b \right) Y_{l_2 -m_2}\left( \hat{e}_b \right) \,.
\end{split}
\end{equation}
The first two lines on the right hand side of (\ref{eq:secondmomenthd_almost}) can be recast as
\begin{equation}
    \sum_{l m} C_l Y^*_{lm}\left(\hat{e}_a\right)Y_{lm}\left(\hat{e}_b\right) \sum_{l' m'} C_{l'} Y^*_{l'm'}\left(\hat{e}_a\right)Y_{l'm'}\left(\hat{e}_b\right) = \left( \gamma_{ab}\left(\zeta\right) \right)^2 \,,
\end{equation}
which is the square of the ORF. Now, by making use of the parity identity of the spherical harmonics,
\begin{equation}
    Y^*_{lm}\left(\hat{e}\right) = (-1)^m Y_{l-m}\left(\hat{e}\right) \,,
\end{equation}
the last sum in (\ref{eq:secondmomenthd_almost}) (third line) can be written as
\begin{equation}
    \sum_{l m} C_l Y^*_{lm}\left(\hat{e}_a\right) Y_{lm}\left(\hat{e}_a\right) \sum_{l'm'} C_{l'} Y_{l'm'}^*\left( \hat{e}_b \right) Y_{l'm'}\left( \hat{e}_b \right) = \left( \gamma_{aa} \right)^2 \,,
\end{equation}
which we identify as the square of the autocorrelation $\gamma_{aa}$.

Simplifying the above expressions leads to (\ref{eq:secondmoment_correlation}), which eventually gets to the total variance (\ref{eq:totalvariance}).

\subsection{Cosmic variance}
\label{subsec:cosmicvar_appendix}

To obtain the cosmic variance (\ref{eq:cv_fullskydef}), we calculate ensemble average of the square of the sky averaged correlation operator.

We start by taking the square of the full sky averaged correlation operator,
\begin{equation}
    \{ {\bm \beta}_a^\dagger {\bm \beta}_b \}_\text{S}^2 = \sum_{ll'} \dfrac{(2l + 1)(2l' + 1)}{(4\pi)^2} {\bm C}_l {\bm C}_{l'} P_l \left( \cos \zeta \right) P_{l'} \left( \cos \zeta \right) \,,
\end{equation}
and getting the ensemble average,
\begin{equation}
    \langle \{ {\bm \beta}_a^\dagger {\bm \beta}_b \}_\text{S}^2 \rangle = \sum_{ll'} \dfrac{(2l + 1)(2l' + 1)}{(4\pi)^2} \langle {\bm C}_l {\bm C}_{l'} \rangle P_l \left( \cos \zeta \right) P_{l'} \left( \cos \zeta \right) \,.
\end{equation}
We then simplify $\langle {\bm C}_l {\bm C}_{l'} \rangle$ by using the definition (\ref{eq:ps_operator}),
\begin{equation}
\begin{split}
    \langle {\bm C}_l {\bm C}_{l'} \rangle = & \bigg \langle \sum_m \dfrac{{\bm b}_{lm}^\dagger {\bm b}_{lm}}{2l + 1} \sum_{m'} \dfrac{{\bm b}_{l'm'}^\dagger {\bm b}_{l'm'}}{2l' + 1} \bigg \rangle \\
    = & \sum_{mm'} \dfrac{1}{(2l+1)(2l'+1)} \langle {\bm b}_{lm}^\dagger {\bm b}_{lm} {\bm b}_{l'm'}^\dagger {\bm b}_{l'm'} \rangle \,,
\end{split}
\end{equation}
and performing the Wick rotation (\ref{eq:wickrotation}), 
\begin{equation}
    \langle {\bm b}_{lm}^\dagger {\bm b}_{lm} {\bm b}_{l'm'}^\dagger {\bm b}_{l'm'} \rangle = C_l C_{l'} + C_l^2 \delta_{ll'} \delta_{mm'} + C_l^2 \delta_{ll'} \delta_{m-m'} \,,
\end{equation}
thereby assuming that the fields involved are statistically Gaussian.

These lead to (\ref{eq:secondmoment_corr_skyave}) and so eventually to (\ref{eq:cv_ps}).

\subsection{Variance in the power spectrum}
\label{subsec:ps_variance_appendix}

To calculate the variance in the power spectrum, we obtain the ensemble average of the second moment of the power spectrum multipoles.

We start by taking the ensemble average of its square,
\begin{equation}
\begin{split}
    \langle {\bm C}_l^2 \rangle = & \bigg \langle \sum_{mm'} \dfrac{{\bm b}_{lm}^\dagger {\bm b}_{lm} {\bm b}_{lm'}^\dagger {\bm b}_{lm'}}{(2l + 1)^2} \bigg \rangle \\
    = & \dfrac{1}{(2l + 1)^2} \sum_{mm'} \langle {\bm b}_{lm}^\dagger {\bm b}_{lm} {\bm b}_{lm'}^\dagger {\bm b}_{lm'} \rangle \,.
\end{split}
\end{equation}
By the Gaussian factorization (\ref{eq:wickrotation}),
\begin{equation}
    \langle {\bm b}_{lm}^\dagger {\bm b}_{lm} {\bm b}_{lm'}^\dagger {\bm b}_{lm'} \rangle = C_l^2 \left( 1 + \delta_{mm'} + \delta_{m-m'} \right) \,,
\end{equation}
we move up the last line to
\begin{equation}
\begin{split}
    \langle {\bm C}_l^2 \rangle = & \dfrac{C_l^2}{(2l + 1)^2} \sum_{mm'} \left( 1 + \delta_{mm'} + \delta_{m-m'} \right) \\
    = & \dfrac{C_l^2}{(2l + 1)^2} \left[ (2l + 1)^2 + 2 (2l + 1) \right] \,.
\end{split}
\end{equation}
We eventually simplify this to
\begin{equation}
\label{eq:psop_secondmoment}
    \langle {\bm C}_l^2 \rangle = C_l^2 \left( 1 + \dfrac{1}{l + (1/2)} \right) \,.
\end{equation}

Substituting the above result into (\ref{eq:psop_variance_def}) leads to (\ref{eq:psop_variance}).


\begin{thebibliography}{10}

\bibitem{LIGOScientific:2016aoc}
{\scshape LIGO Scientific, Virgo} collaboration, \emph{{Observation of
  Gravitational Waves from a Binary Black Hole Merger}},
  \href{https://doi.org/10.1103/PhysRevLett.116.061102}{\emph{Phys. Rev. Lett.}
  {\bfseries 116} (2016) 061102}
  [\href{https://arxiv.org/abs/1602.03837}{{\ttfamily 1602.03837}}].

\bibitem{LIGOScientific:2017vwq}
{\scshape LIGO Scientific, Virgo} collaboration, \emph{{GW170817: Observation
  of Gravitational Waves from a Binary Neutron Star Inspiral}},
  \href{https://doi.org/10.1103/PhysRevLett.119.161101}{\emph{Phys. Rev. Lett.}
  {\bfseries 119} (2017) 161101}
  [\href{https://arxiv.org/abs/1710.05832}{{\ttfamily 1710.05832}}].

\bibitem{LIGOScientific:2021sio}
{\scshape LIGO Scientific, VIRGO, KAGRA} collaboration, \emph{{Tests of General
  Relativity with GWTC-3}},  \href{https://arxiv.org/abs/2112.06861}{{\ttfamily
  2112.06861}}.

\bibitem{LIGOScientific:2021djp}
{\scshape LIGO Scientific, VIRGO, KAGRA} collaboration, \emph{{GWTC-3: Compact
  Binary Coalescences Observed by LIGO and Virgo During the Second Part of the
  Third Observing Run}},  \href{https://arxiv.org/abs/2111.03606}{{\ttfamily
  2111.03606}}.

\bibitem{KAGRA:2018plz}
{\scshape KAGRA} collaboration, \emph{{KAGRA: 2.5 Generation Interferometric
  Gravitational Wave Detector}},
  \href{https://doi.org/10.1038/s41550-018-0658-y}{\emph{Nature Astron.}
  {\bfseries 3} (2019) 35} [\href{https://arxiv.org/abs/1811.08079}{{\ttfamily
  1811.08079}}].

\bibitem{LISA:2022kgy}
{\scshape LISA} collaboration, \emph{{New Horizons for Fundamental Physics with
  LISA}},  \href{https://arxiv.org/abs/2205.01597}{{\ttfamily 2205.01597}}.

\bibitem{TianQin:2015yph}
{\scshape TianQin} collaboration, \emph{{TianQin: a space-borne gravitational
  wave detector}},
  \href{https://doi.org/10.1088/0264-9381/33/3/035010}{\emph{Class. Quant.
  Grav.} {\bfseries 33} (2016) 035010}
  [\href{https://arxiv.org/abs/1512.02076}{{\ttfamily 1512.02076}}].

\bibitem{Detweiler:1979wn}
S.~L. Detweiler, \emph{{Pulsar timing measurements and the search for
  gravitational waves}}, \href{https://doi.org/10.1086/157593}{\emph{Astrophys.
  J.} {\bfseries 234} (1979) 1100}.

\bibitem{Burke-Spolaor:2018bvk}
S.~Burke-Spolaor et~al., \emph{{The Astrophysics of Nanohertz Gravitational
  Waves}}, \href{https://doi.org/10.1007/s00159-019-0115-7}{\emph{Astron.
  Astrophys. Rev.} {\bfseries 27} (2019) 5}
  [\href{https://arxiv.org/abs/1811.08826}{{\ttfamily 1811.08826}}].

\bibitem{Hellings:1983fr}
R.~W. Hellings and G.~W. Downs, \emph{{Upper limits on the isotropic
  gravitational radiation background from pulsar timing analysis}},
  \href{https://doi.org/10.1086/183954}{\emph{Astrophys. J. Lett.} {\bfseries
  265} (1983) L39}.

\bibitem{Romano:2019yrj}
J.~D. Romano, \emph{{Searches for stochastic gravitational-wave backgrounds}},
  8, 2019, \href{https://arxiv.org/abs/1909.00269}{{\ttfamily 1909.00269}}.

\bibitem{NANOGrav:2020bcs}
{\scshape NANOGrav} collaboration, \emph{{The NANOGrav 12.5 yr Data Set: Search
  for an Isotropic Stochastic Gravitational-wave Background}},
  \href{https://doi.org/10.3847/2041-8213/abd401}{\emph{Astrophys. J. Lett.}
  {\bfseries 905} (2020) L34}
  [\href{https://arxiv.org/abs/2009.04496}{{\ttfamily 2009.04496}}].

\bibitem{Shannon:2015ect}
R.~M. Shannon et~al., \emph{{Gravitational waves from binary supermassive black
  holes missing in pulsar observations}},
  \href{https://doi.org/10.1126/science.aab1910}{\emph{Science} {\bfseries 349}
  (2015) 1522} [\href{https://arxiv.org/abs/1509.07320}{{\ttfamily
  1509.07320}}].

\bibitem{Lentati:2015qwp}
L.~Lentati et~al., \emph{{European Pulsar Timing Array Limits On An Isotropic
  Stochastic Gravitational-Wave Background}},
  \href{https://doi.org/10.1093/mnras/stv1538}{\emph{Mon. Not. Roy. Astron.
  Soc.} {\bfseries 453} (2015) 2576}
  [\href{https://arxiv.org/abs/1504.03692}{{\ttfamily 1504.03692}}].

\bibitem{2010CQGra..27h4013H}
G.~{Hobbs}
  et~al.\href{https://doi.org/10.1088/0264-9381/27/8/084013}{\emph{Classical
  and Quantum Gravity} {\bfseries 27} (2010) 084013}
  [\href{https://arxiv.org/abs/0911.5206}{{\ttfamily 0911.5206}}].

\bibitem{Allen:2022dzg}
B.~Allen, \emph{{Variance of the Hellings-Downs Correlation}},
  \href{https://arxiv.org/abs/2205.05637}{{\ttfamily 2205.05637}}.

\bibitem{Allen:2022ksj}
B.~Allen and J.~D. Romano, \emph{{The Hellings and Downs correlation of an
  arbitrary set of pulsars}},
  \href{https://arxiv.org/abs/2208.07230}{{\ttfamily 2208.07230}}.

\bibitem{Qin:2018yhy}
W.~Qin, K.~K. Boddy, M.~Kamionkowski and L.~Dai, \emph{{Pulsar-timing arrays,
  astrometry, and gravitational waves}},
  \href{https://doi.org/10.1103/PhysRevD.99.063002}{\emph{Phys. Rev. D}
  {\bfseries 99} (2019) 063002}
  [\href{https://arxiv.org/abs/1810.02369}{{\ttfamily 1810.02369}}].

\bibitem{Qin:2020hfy}
W.~Qin, K.~K. Boddy and M.~Kamionkowski, \emph{{Subluminal stochastic
  gravitational waves in pulsar-timing arrays and astrometry}},
  \href{https://doi.org/10.1103/PhysRevD.103.024045}{\emph{Phys. Rev. D}
  {\bfseries 103} (2021) 024045}
  [\href{https://arxiv.org/abs/2007.11009}{{\ttfamily 2007.11009}}].

\bibitem{Ng:2021waj}
K.-W. Ng, \emph{{Redshift-space fluctuations in stochastic gravitational wave
  background}}, \href{https://doi.org/10.1103/PhysRevD.106.043505}{\emph{Phys.
  Rev. D} {\bfseries 106} (2022) 043505}
  [\href{https://arxiv.org/abs/2106.12843}{{\ttfamily 2106.12843}}].

\bibitem{Liu:2022skj}
G.-C. Liu and K.-W. Ng, \emph{{Timing-residual power spectrum of a polarized
  stochastic gravitational-wave background in pulsar-timing-array
  observation}}, \href{https://doi.org/10.1103/PhysRevD.106.064004}{\emph{Phys.
  Rev. D} {\bfseries 106} (2022) 064004}
  [\href{https://arxiv.org/abs/2201.06767}{{\ttfamily 2201.06767}}].

\bibitem{Bernardo:2022rif}
R.~C. Bernardo and K.-W. Ng, \emph{{Stochastic gravitational wave background
  phenomenology in a pulsar timing array}},
  \href{https://arxiv.org/abs/2208.12538}{{\ttfamily 2208.12538}}.

\bibitem{10.5555/1593511}
G.~Van~Rossum and F.~L. Drake, \emph{Python 3 Reference Manual}. CreateSpace,
  Scotts Valley, CA, 2009.

\bibitem{Julia-2017}
J.~Bezanson, A.~Edelman, S.~Karpinski and V.~B. Shah, \emph{Julia: A fresh
  approach to numerical computing},
  \href{https://doi.org/10.1137/141000671}{\emph{SIAM {R}eview} {\bfseries 59}
  (2017) 65}.

\bibitem{1998IJMPD...7...89N}
K.-W. {Ng}, \emph{{Sample Variance in Large-Scale Cosmic Microwave Background
  Anisotropy Measurements}},
  \href{https://doi.org/10.1142/S0218271898000097}{\emph{Int. J. Mod. Phys. D}
  {\bfseries 7} (1998) 89}.

\bibitem{Ng:1997ez}
K.-W. Ng and G.-C. Liu, \emph{{Correlation functions of CMB anisotropy and
  polarization}}, \href{https://doi.org/10.1142/S0218271899000079}{\emph{Int.
  J. Mod. Phys. D} {\bfseries 8} (1999) 61}
  [\href{https://arxiv.org/abs/astro-ph/9710012}{{\ttfamily
  astro-ph/9710012}}].

\bibitem{Gair:2014rwa}
J.~Gair, J.~D. Romano, S.~Taylor and C.~M.~F. Mingarelli, \emph{{Mapping
  gravitational-wave backgrounds using methods from CMB analysis: Application
  to pulsar timing arrays}},
  \href{https://doi.org/10.1103/PhysRevD.90.082001}{\emph{Phys. Rev. D}
  {\bfseries 90} (2014) 082001}
  [\href{https://arxiv.org/abs/1406.4664}{{\ttfamily 1406.4664}}].

\bibitem{NANOGrav:2021ini}
{\scshape NANOGrav} collaboration, \emph{{The NANOGrav 12.5-year Data Set:
  Search for Non-Einsteinian Polarization Modes in the Gravitational-wave
  Background}},
  \href{https://doi.org/10.3847/2041-8213/ac401c}{\emph{Astrophys. J. Lett.}
  {\bfseries 923} (2021) L22}
  [\href{https://arxiv.org/abs/2109.14706}{{\ttfamily 2109.14706}}].

\bibitem{deRham:2018red}
C.~de~Rham and S.~Melville, \emph{{Gravitational Rainbows: LIGO and Dark Energy
  at its Cutoff}},
  \href{https://doi.org/10.1103/PhysRevLett.121.221101}{\emph{Phys. Rev. Lett.}
  {\bfseries 121} (2018) 221101}
  [\href{https://arxiv.org/abs/1806.09417}{{\ttfamily 1806.09417}}].

\bibitem{Liang:2021bct}
Q.~Liang and M.~Trodden, \emph{{Detecting the stochastic gravitational wave
  background from massive gravity with pulsar timing arrays}},
  \href{https://doi.org/10.1103/PhysRevD.104.084052}{\emph{Phys. Rev. D}
  {\bfseries 104} (2021) 084052}
  [\href{https://arxiv.org/abs/2108.05344}{{\ttfamily 2108.05344}}].

\bibitem{Tachinami:2021jnf}
T.~Tachinami, S.~Tonosaki and Y.~Sendouda, \emph{{Gravitational-wave
  polarizations in generic linear massive gravity and generic higher-curvature
  gravity}}, \href{https://doi.org/10.1103/PhysRevD.103.104037}{\emph{Phys.
  Rev. D} {\bfseries 103} (2021) 104037}
  [\href{https://arxiv.org/abs/2102.05540}{{\ttfamily 2102.05540}}].

\bibitem{Bernardo:2022vlj}
R.~C. Bernardo and K.-W. Ng, \emph{{Looking out for the Galileon in the
  nanohertz gravitational wave sky}},
  \href{https://arxiv.org/abs/2206.01056}{{\ttfamily 2206.01056}}.

\bibitem{Smits:2008cf}
R.~Smits, M.~Kramer, B.~Stappers, D.~R. Lorimer, J.~Cordes and A.~Faulkner,
  \emph{{Pulsar searches and timing with the square kilometre array}},
  \href{https://doi.org/10.1051/0004-6361:200810383}{\emph{Astron. Astrophys.}
  {\bfseries 493} (2009) 1161}
  [\href{https://arxiv.org/abs/0811.0211}{{\ttfamily 0811.0211}}].

\bibitem{Lazio:2013mea}
T.~J.~W. Lazio, \emph{{The Square Kilometre Array pulsar timing array}},
  \href{https://doi.org/10.1088/0264-9381/30/22/224011}{\emph{Class. Quant.
  Grav.} {\bfseries 30} (2013) 224011}.

\bibitem{Janssen:2014dka}
G.~Janssen et~al., \emph{{Gravitational wave astronomy with the SKA}},
  \href{https://doi.org/10.22323/1.215.0037}{\emph{PoS} {\bfseries AASKA14}
  (2015) 037} [\href{https://arxiv.org/abs/1501.00127}{{\ttfamily
  1501.00127}}].

\bibitem{Miao:2021awa}
X.~Miao, H.~Xu, L.~Shao, C.~Liu and B.-Q. Ma, \emph{{Stringent Tests of Gravity
  with Highly Relativistic Binary Pulsars in the Era of LISA and SKA}},
  \href{https://doi.org/10.3847/1538-4357/ac1d48}{\emph{Astrophys. J.}
  {\bfseries 921} (2021) 114}
  [\href{https://arxiv.org/abs/2107.05812}{{\ttfamily 2107.05812}}].

\bibitem{Joshi:2022ohi}
B.~C. Joshi et~al., \emph{{Nanohertz Gravitational Wave Astronomy during the
  SKA Era: An InPTA perspective}},
  \href{https://arxiv.org/abs/2207.06461}{{\ttfamily 2207.06461}}.

\bibitem{Bustamante-Rosell:2021daj}
M.~J. Bustamante-Rosell, J.~Meyers, N.~Pearson, C.~Trendafilova and
  A.~Zimmerman, \emph{{Gravitational wave timing array}},
  \href{https://doi.org/10.1103/PhysRevD.105.044005}{\emph{Phys. Rev. D}
  {\bfseries 105} (2022) 044005}
  [\href{https://arxiv.org/abs/2107.02788}{{\ttfamily 2107.02788}}].

\end{thebibliography}

\providecommand{\href}[2]{#2}\begingroup\raggedright\endgroup

\end{document}